\documentclass[10pt,letterpaper]{article}
\usepackage[top=0.85in,left=2.75in,footskip=0.75in]{geometry}

\usepackage{changepage}

\usepackage[utf8]{inputenc}

\usepackage{textcomp,marvosym}

\usepackage{fixltx2e}

\usepackage{amsmath,amssymb}

\usepackage{cite}

\usepackage{nameref,hyperref}

\usepackage{microtype}
\DisableLigatures[f]{encoding = *, family = * }

\usepackage{rotating}

\raggedright
\usepackage[aboveskip=1pt,labelfont=bf,labelsep=period,justification=raggedright,singlelinecheck=off]{caption}

\makeatletter
\renewcommand{\@biblabel}[1]{\quad#1.}
\makeatother

\date{}

\usepackage{lastpage,fancyhdr,graphicx}
\usepackage{epstopdf}
\fancyheadoffset[L]{2.25in}
\fancyfootoffset[L]{2.25in}
\usepackage{multirow}

\newcommand{\bcat}[0]{$\beta$-catenin}

\newcommand{\gsk}[0]{GSK$3\beta$}
\newcommand{\hes}[0]{\emph{Hes1}}
\newcommand{\hath}[0]{\emph{Hath1}}

\newcommand{\invitro}[0]{\emph{in vitro}}
\newcommand{\invivo}[0]{\emph{in vivo}}

\newcommand{\etal}[0]{\emph{et al.}}

\usepackage{framed}
\usepackage[mathscr]{euscript}
\usepackage{amsthm}
\newtheorem{mydef}{Definition}

\newtheorem{myprop}{Proposition}
\newtheorem{myremark}{Remark}
\usepackage[version=3]{mhchem}

\usepackage{pifont}
\newcommand{\tick}{\ding{52}}
\newcommand{\cross}{\ding{56}}

\newcommand{\matlab}[0]{\textsc{Matlab}}

\usepackage{tikz}
\usepackage{enumitem}
\newcommand*\mycirc[1]{%
\begin{tikzpicture}[baseline=(C.base)]
\node[draw,circle,inner sep=1pt,minimum size=3ex](C) {#1};
\end{tikzpicture}}

\begin{document}
\vspace*{0.35in}

\begin{flushleft}
{\Large
\textbf\newline{The Role of the Hes1 Crosstalk Hub in Notch-Wnt Interactions of the Intestinal Crypt}
}
\newline
\\
Sophie K. Kay*\textsuperscript{1},
Heather A. Harrington\textsuperscript{2},
Sarah Shepherd\dag,
Keith Brennan\textsuperscript{3},
Trevor Dale\textsuperscript{4},
James M. Osborne\textsuperscript{5},
David J. Gavaghan\textsuperscript{1},
Helen M. Byrne**\textsuperscript{1,2},
\\
\bigskip
\bf{1} Department of Computer Science, University of Oxford, Oxford, U.K.
\\
\bf{2} Mathematical Institute, University of Oxford, Oxford, U.K.
\\
\bf{3} Wellcome Trust Centre for Cell Matrix Research, University of Manchester, Manchester, U.K.
\\
\bf{4} School of Biosciences, Cardiff University, Cardiff, U.K.
\\
\bf{5} School of Mathematics and Statistics, University of Melbourne, Melbourne, Australia.
\\
\dag Deceased
\\
\bigskip

%
%




* sophie.kay@cs.ox.ac.uk \\
** helen.byrne@maths.ox.ac.uk

\end{flushleft}
\section*{Abstract}
The Notch pathway plays a vital role in determining whether cells in the intestinal epithelium adopt a secretory or an absorptive phenotype. Cell fate specification is coordinated via Notch's interaction with the canonical Wnt pathway. Here, we propose a new mathematical model of the Notch and Wnt pathways, in which the Hes1 promoter acts as a hub for pathway crosstalk. Computational simulations of the model can assist in understanding how healthy intestinal tissue is maintained, and predict the likely consequences of biochemical knockouts upon cell fate selection processes. Chemical reaction network theory (CRNT) is a powerful, generalised framework which assesses the capacity of our model for monostability or multistability, by analysing properties of the underlying network structure without recourse to specific parameter values or functional forms for reaction rates. CRNT highlights the role of \bcat{} in stabilising the Notch pathway and damping oscillations, demonstrating that Wnt-mediated actions on the Hes1 promoter can induce dynamical transitions in the Notch system, from multistability to monostability. Time-dependent model simulations of cell pairs reveal the stabilising influence of Wnt upon the Notch pathway, in which \bcat- and Dsh-mediated action on the Hes1 promoter are key in shaping the subcellular dynamics. Where Notch-mediated transcription of Hes1 dominates, there is Notch oscillation and maintenance of fate flexibility; Wnt-mediated transcription of Hes1 favours bistability akin to cell fate selection. Cells could therefore regulate the proportion of Wnt- and Notch-mediated control of the Hes1 promoter to coordinate the timing of cell fate selection as they migrate through the intestinal epithelium and are subject to reduced Wnt stimuli. Furthermore, mutant cells characterised by hyperstimulation of the Wnt pathway may, through coupling with Notch, invert cell fate in neighbouring healthy cells, enabling an aberrant cell to maintain its neighbours in mitotically active states.

\section*{Author Summary}
Epithelial cells which line the intestine form finger-shaped structures called crypts; these undergo a process of renewal at the base, causing cells to migrate upwards until they die and are sloughed off into the gut. Much of our understanding of how crypts function rests upon two processes: proliferation, in which cells divide to produce `daughter cells'; and differentiation, in which cells become progressively more specialised as they migrate along the crypt axis and mature. Coordinated proliferation and differentiation enable the crypt to renew itself and to produce a range of specialised cell types essential to its healthy functioning. In this paper we build a mathematical model for two reaction pathways which regulate proliferation and differentiation. We use this model to explore how crosstalk between these pathways in cell pairs influences the generation of distinct cell fates in intestinal tissues. By modifying our model to represent abnormal, `mutant' cells, we investigate abnormalities typical of early colorectal cancer. Computational simulation of our model identifies an important region of crosstalk in our reaction network which determines whether cells adopt the same fate as one another, or different fates. Our model may prove useful for realistic simulations of whole crypts in the future.


\section*{Introduction}
Attainment and maintenance of homeostasis within the epithelial lining of the intestine is achieved through a nuanced coordination of biochemical processes and spatial cues within the tissue, in particular those which influence proliferation and cell fate selection. Crosstalk between the subcellular pathways governing these processes facilitates the coordination of cellular division and specialisation throughout the tissue of the intestinal epithelium, producing the broad range of cell types required for its function, ranging from totipotent stem cells, to terminally differentiated secretory or absorptive phenotypes. These cells collectively form test-tube shaped structures called \emph{crypts}, each consisting of up to $700$ cells in mice \cite{rodriguez1979cell} and 2000 in humans \cite{nicolas2007stem}, with millions of such crypts distributed throughout the intestine. As shown in the schematics of Figs. \ref{fig: theta2}A and \ref{fig: mutant-hits}A, cells proliferate in the lower regions of the crypt \cite{quastler1959cell} and (with the exception of specialised Paneth cells \cite{porter2002multifaceted}) migrate upwards, until they die and are sloughed into the lumen of the gut \cite{potten1997role}. Increased specialisation occurs as cells migrate up the crypt, arguably due to spatial cues in the surrounding tissue \cite{sato2009single, ootani2009sustained}. Consequently the interlacing of proliferative and differential processes is key to understanding the attainment and maintenance of homeostasis in the healthy intestinal epithelium, or indeed how this is perturbed in conditions such as colorectal cancer.

In this paper we focus on interactions between the canonical Wnt pathway, in its role of mitotic regulator, and the Notch pathway, in its role of cell fate specifier. Both pathways facilitate responses to spatial cues within the tissue: the canonical Wnt pathway characterises the cellular response to local extracellular concentrations of Wnt, whilst the Notch pathway coordinates cell fate progression in populations of neighbouring cells.

The \emph{Wnt pathway} is crucial in the development and maintenance of biological tissues. In its canonical form, \bcat{} assists in regulating transitions through the cell cycle and centres on a mechanism for the control of cytoplasmic levels of \bcat{}. When extracellular Wnt stimulus is low, \bcat{} degradation is elevated; translocation of \bcat{} to the nucleus is limited and the production of target genes is reduced. In the presence of a high Wnt stimulus, there is an increase in cytoplasmic levels of \bcat{} and, consequently, increased translocation of \bcat{} to the nucleus and production of associated target genes, including some associated with cell-cycle progression \cite{davidson2010emerging}. Of further relevance within the context of the intestinal crypt is the spatial variation in extracellular Wnt along the  crypt axis, with a high concentration at the crypt base which tapers off towards the crypt mouth \cite{gregorieff2005expression}. Non-canonical forms of Wnt signalling govern a variety of processes, including: integrin-mediated intercellular adhesion; planar cell polarity, and Wnt/calcium signalling \cite{lustig2003wnt}. In this study, we focus on the canonical pathway (discussion of other modes of Wnt action can be found in the reviews \cite{amerongen2012alternative, korswagen2002canonical, strutt2003frizzled, veeman2003second}).

A number of mathematical models of the canonical Wnt pathway have been proposed, typically using ordinary differential equations (ODEs). Of particular note is the ODE model of Lee \etal{} \cite{lee2003roles}, which proposes that Axin, a component of the \bcat{} destruction complex, acts as a rate-limiter in \bcat{} degradation and may provide a robust regulator for the strength of Wnt signalling within a cell. Model simulations \cite{tolwinski2004rethinking, lee2003roles} have led to suggestions that Axin might  regulate crosstalk between Wnt and other pathways (e.g. MAPK), by stabilising levels of other members of the destruction complex such as APC. Systematic analyses have yielded reduced ODE systems \cite{kruger2004model, mirams2010multiple, lloyd2013toward}, have reformulated the model as a system of delay differential equations \cite{wawra2007extended}, have added inhibitory Wnt targets to generate oscillations \cite{cho2006wnt, kim2007hidden}, or have extended the network to resolve the \bcat{} degradation processes in greater detail \cite{benary2015mathematical}. Computational evaluation \cite{goentoro2009evidence} of the model of Lee \etal{} found that the fold-change in, rather than the absolute expression of, \bcat{} was the most robust feature to parameter perturbations, given a fixed Wnt stimulus. This fold-change may act as a means of overcoming biological noise and is observed experimentally in human colorectal cell lines \cite{goentoro2009evidence}. Recalibration of the model of Lee \etal{} for several human and canine cell lines \cite{tan2012wnt} suggests that higher Axin and lower APC concentrations are required when applying the model in a mammalian context, rather than the \emph{Xenopus} oocytes of Lee \etal. Alternative mathematical models of the Wnt pathway adopt shuttling and/or compartmental approaches, accounting for subcellular localisation of proteins in the nucleus, cytoplasm and cell membrane \cite{leeuwen2007elucidating, schmitz2011nucleo, schmitz2013analysing, maclean2015parameter, gross2016algebraic}.

The \emph{Notch pathway} regulates the transition from fully undifferentiated to terminally differentiated cell and belongs to the class of \emph{juxtacrine} signalling networks \cite{artavanis1999notch}. Juxtacrine signals are initiated by contact-based processes, in which two or more adjacent cells transfer signals via ligand binding events at their cell surface membranes \cite{massague1990transforming}. In the intestinal epithelium, the Notch pathway is associated with the generation of \emph{salt-and-pepper} patterns involving secretory and absorptive cell phenotypes. High Notch activity marks a cell for conversion to absorptive type, whereas cells with low Notch activity ultimately become secretory. Moreover, there are four types of differentiated cell in the epithelium of the small intestine: goblet, Paneth, and enteroendocrine, which are secretory; and enterocytes, which are absorptive \cite{bjerknes1999clonal, yeung2011regulation}. The Notch pathway plays an important role in selecting which cell fate will be realised, via the Notch target \hes, and the protein \hath, which is normally suppressed by \hes. Expression of \hath{} (and suppression of \hes) is associated with secretory phenotypes \cite{yang2001requirement}, whilst high \hes{} expression correlates with an absorptive fate. \emph{In vivo} studies on the small intestine of \hes-knockout mice have generated substantial numbers of all three secretory cell types \cite{jensen2000control}. Similarly, inhibition of Notch signalling yields an increased population of goblet cells \cite{es2005notch}. Gain- and loss-of-function studies by Fre \etal{} \cite{fre2005notch} support this theory and suggest that Notch signalling via \hes{} is responsible for early-stage cell fate selection. As a cell migrates through the epithelium and is exposed to reduced Wnt levels, other pathways and cell regulators coordinate a more refined selection of fate: they may, for example, determine a specific fate for a cell which has already been selected for secretory function \cite{yeung2011regulation, nakamura2007crosstalk}. We conclude that understanding how \hes{} and \hath{} expression levels are regulated within a cell may offer valuable insights into the cell fate selection process.

Many mathematical models of Notch signalling focus on receptor-ligand binding, using either discrete difference equations or continuous ODEs. Such models have captured lateral inhibition \cite{collier1996pattern, wearing2000mathematical} or lateral induction \cite{owen1998mathematical, owen2000lateral, webb2004oscillations} and may be embedded in individual cells of a lattice to generate cell patterns. Other models distinguish between cis- and trans-Delta in the Notch binding events, in which Notch may bind to either a \emph{cis-}Delta ligand on the same cell, or a \emph{trans-}Delta ligand on a neighbouring cell; this generates strong switch-like behaviour \cite{sprinzak2010cis}. Many mathematical models have included subcellular detail of the Notch pathway, typically building on the Goodwin model for negative feedback, which employs ordinary differential equations and augments a system of mRNA and protein with an unspecified intermediate in order to generate sustained oscillations \cite{goodwin1965oscillatory}. Hirata \etal{} \cite{hirata2002oscillatory} follow Goodwin and propose a simple ODE model which achieves an experimentally-validated, two hour oscillation period for \hes{}. Other reformulations of the Goodwin framework highlight the functional importance of transcriptional delays in modelling oscillations in Hes1 mRNA and protein levels \cite{monk2003oscillatory, lewis2003autoinhibition}, with dimerisation of \hes{} protein yielding oscillation amplitudes in closer agreement with experimental data \cite{momiji2008dissecting}. Autorepression of \hes{} has been shown to aid the tunability of oscillations when two \hes{} oscillators are coupled \cite{momiji2009oscillatory}. Detailed subcellular models suggest that variation in transcriptional repression of \hes{} may facilitate the transition between oscillation and bistability \cite{agrawal2009computational} and may be used to generate early cell-fate selection \cite{kiparissides2011modelling}, although crosstalk with another pathway is required for terminal differentiation \cite{kiparissides2011modelling}.

Simple mathematical and computational models for Notch-Wnt interaction are also employed in multicellular settings to study the effects of spatial variation in Wnt stimuli upon cell fate selection. Some of these models eschew equation systems altogether, in favour of a rule-based paradigm in which Notch and Wnt activity are each either high or low and the system evolves according to an averaging process over neighbouring cells \cite{buske2011comprehensive, pin2012modelling}. However, this pared-down approach cannot represent subcellular details of the crosstalk. Existing mathematical models at the subcellular scale have accounted for interactions between \gsk{} and NICD \cite{goldbeter2008modeling}, between NICD and Dsh \cite{wang2013mathematical} or via LEF/TCF and membrane-bound Notch, in which inhibition of Wnt is used to drive terminal differentiation \cite{kirnasovsky2008analysis, agur2011dickkopf1}. Some of these models omit a \hes{} autoregulation motif \cite{kirnasovsky2008analysis, agur2011dickkopf1} and indeed, few multicellular studies of the role of \hes{} oscillations in cell fate exist at the present time. Oscillations of \hes{} are thought to assist in the maintenance of cell-cycle processes and the flexibility of cell fate decisions within a cell population \cite{kageyama2008dynamic}.

Our main focus is therefore the development of a mathematical model for the interaction of the Notch and Wnt pathways in the cells of the intestinal epithelium. Given the rich complexity which reaction networks and their crosstalk generate, computational approaches form a cornerstone of the analysis and implementation of our model. In particular, chemical reaction network theory (CRNT) offers a powerful means of analysing the steady-state behaviour of our model without recourse to either specific parameter values or functional forms for the reaction dynamics. CRNT facilitates the analysis of large and complex networks and is performed with a ready-to-use computational toolbox \cite{ji2013crnt}. Having derived, parametrised and calibrated our model, we use it to explore two-cell systems in healthy and malignant scenarios. These cell pair settings demonstrate the main dynamical features of the model, illustrate the role of Notch-Wnt crosstalk in shaping the cell fate response, and permit preliminary investigation of the possible consequences of dysregulation of components of the two pathways in the subcellular network.

\section*{Results}
\begin{figure*}[h]
\centering
\includegraphics[width=\textwidth]{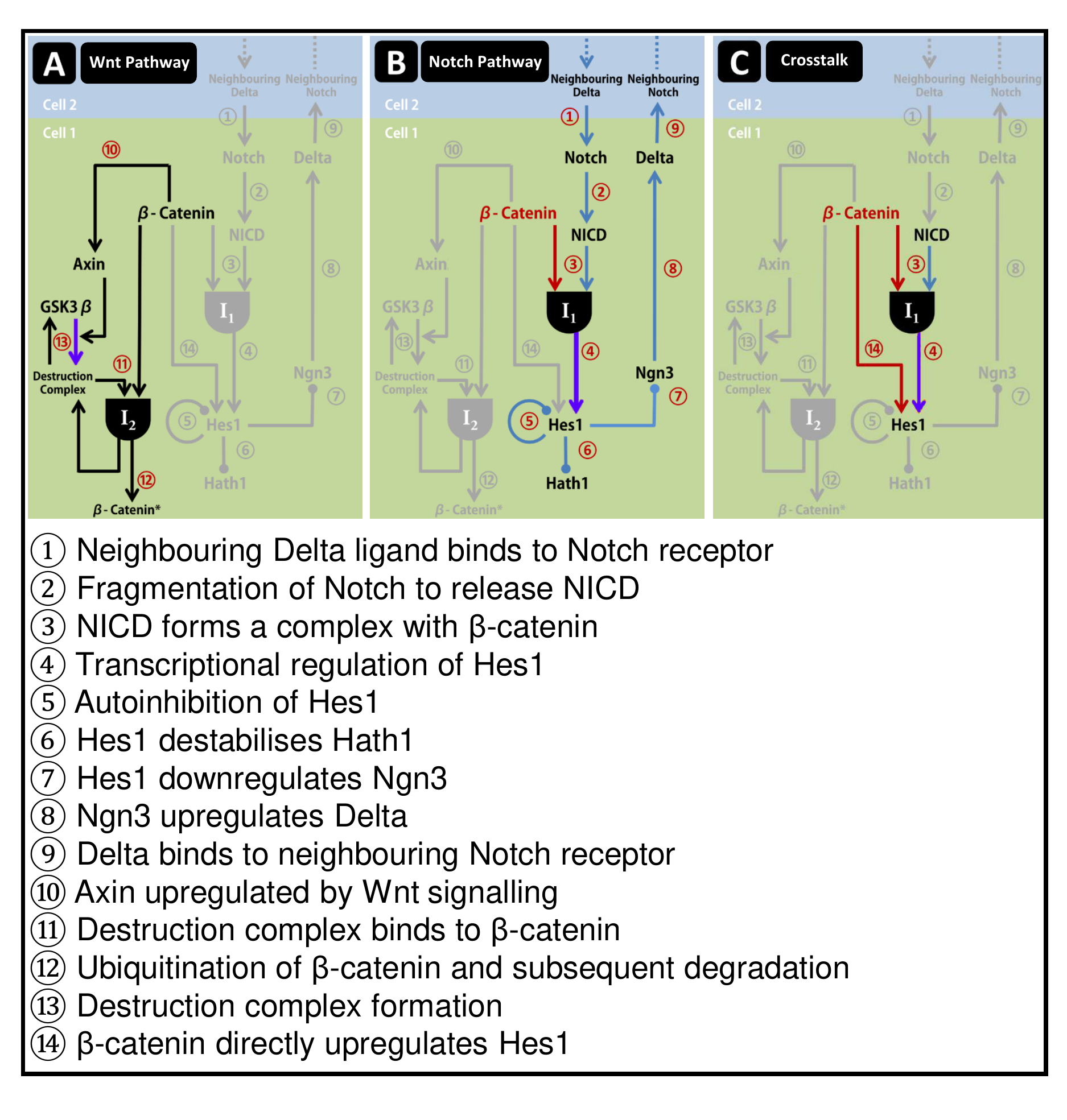}
\caption[]{{\bf Network representation for our model of Notch-Wnt interaction.} Our model comprises: (A) the Wnt pathway, (B) the Notch pathway, and (C) crosstalk points. Numbered steps are justified in the SI. Major steps involving the \bcat{} crosstalk hub are shown in red, while Wnt-dependent steps are shown in purple. Black AND gates signify the formation of intermediate complexes from two molecular partners. Circular end caps indicate inhibition steps.}
\label{fig: NetworkDiagram}
\end{figure*}

The primary aim of our mathematical model is to study how crosstalk between the Notch and Wnt pathways -- in particular activity via the Hes1 promoter -- influences cell fate selection. A schematic of the biochemical network which our model represents is shown in Fig.~\ref{fig: NetworkDiagram}. Each of the $14$ steps in our network model is supported by experimental evidence (references for each step are shown in the SI).

\subsection*{Derivation of Notch-Wnt crosstalk model}
The Wnt elements of our model (five species) represent the shuttling of \gsk{} between a non-complexed form ($G$) and a complexed form ($C$), and their effects upon \bcat{} ($B$) and Axin levels ($A$) (Fig.~\ref{fig: NetworkDiagram}A). We refer to $C$ as the `destruction complex' (comprising \gsk, Axin and - implicitly - APC) because of its role in targeting \bcat{} for ubiquitination and degradation, although we do not explicitly represent these two processes in our model. Instead we consider a complex $I_2$, which comprises the destruction complex bound to \bcat{} and which degrades to release only $C$ back into the system.

The components of our Notch submodel (seven species) encapsulate the binding of Delta ligand ($D$) and Notch receptor ($N$) at the cell surface membrane \cite{vassin1987neurogenic} and the resulting subcellular signalling cascade \cite{artavanis1999notch} responsible for Hes1 regulation (Fig.~\ref{fig: NetworkDiagram}B). Given our focus on crosstalk with the Wnt pathway, our Notch submodel incorporates NICD ($F$) arising from Notch cleavage; an intermediate complex ($I_1$) formed when NICD binds to \bcat; the cell fate specifiers Hes1 ($H_1$) and Hath1 ($H_2$); and the proneural protein Ngn3 ($P$).

Our focus on Notch-Wnt interaction motivates us to consider the crosstalk hub involved with Hes1 regulation, (Fig.~\ref{fig: NetworkDiagram}C). Here, Hes1 regulation is governed by two main routes. The \emph{Notch-mediated} route (Steps \mycirc{3} and \mycirc{4}) involves the binding of $I_1$ to the promoter. This interaction has been identified in vascular progenitor cells \cite{yamamizu2010convergence} and human kidney cells \cite{jin2009beta} and may provide a switching mechanism in the canonical Wnt pathway by diverting \bcat{} from regulating Wnt target genes \cite{jin2009beta}. In the \emph{Wnt-mediated} route, regulation occurs via \bcat{} binding alone (Step \mycirc{14}); this Notch-independent mechanism is supported by Peignon \etal{} \cite{peignon2011complex}, who have identified complementary binding sites on the \bcat{} molecule and Hes1 promoter and infer direct regulation of Hes1 levels by \bcat.

We also incorporate a Wnt-mediated intervention involving the downregulation of Hes1 by Dishevelled \cite{collu2012dishevelled}. Experimental evidence \cite{collu2012dishevelled} reveals a $0.4-$ to $0.5-$fold change in Notch activity in response to expression of either Wnt or Dsh, yet expression of \bcat{} results in a $1.2-$fold change. This interaction is thought to occur through binding and reduction in the levels of a NICD coactivator, CSL, and is distinct from the direct, `Notch-mediated' regulation mechanism described above. For simplicity, we do not model the concentration of Dsh, but we scale the production of Hes1 by a function of the local Wnt stimulus to simulate this effect (full details in SI).

The interactions which comprise our network model are based upon the following assumptions:
\begin{itemize}
 \item Decay rates follow a first-order mass action law (species $X$ decays at rate $\mu_X X$ for $\mu_X$ the rate constant of decay);
 \item Binding reactions between \bcat{} and NICD, and between \bcat{} and the destruction complex, follow first-order mass action laws;
 \item The rate of upregulation events in the Notch system, and of Axin in the Wnt system, are modelled using Hill functions;
 \item Inhibition events are represented by hyperbolas and include the Dsh-mediated downregulation of Hes1, denoted by a Wnt-dependent function $\Psi_W$;
 \item Incorporation of non-complexed \gsk{} into the destruction complex ($G \rightarrow C$) is represented by a function $\Psi_{W,A}$, which depends on the subcellular Axin concentration $A$ and the extracellular Wnt concentration $W$; the reverse reaction ($C \rightarrow G$) is assumed to occur at a constant rate.
\end{itemize}
Under these assumptions, we realise the reaction network as a system of twelve ordinary differential equations (ODEs), of the form $\dot{\bf{x}} = Y(\bf{x},\bf{k})$, where $\bf{x}$ is a vector of the twelve network species, $\dot{\bf{x}}$ represents the time derivative of the species concentrations, and the vector $\bf{k}$ contains the model parameters (see Tables \ref{tab: dimensional-decay-rates} -- \ref{tab: general-rate-params}). The full set of model equations is detailed in the SI, Eqns. (\ref{eqn: notch}) -- (\ref{eqn: axin}).

\subsection*{Steady-state analysis}
The steady state behaviour of the mathematical model is used to confirm qualitative matching with known biological features. Given the size of the system ($12$ variables and $41$ parameters, Table \ref{tab: parameters}), this analysis is restricted to a `single cell' scenario, commensurate with a homogeneous population of cells, in which the Notch and Wnt subnetworks decouple. Terms for NICD in the Wnt system, or \bcat{} in the Notch system, are treated as input parameters rather than variables. This renders  tractable analysis of each subnetwork (see SI for full details and figures).

The Wnt submodel (Fig.~\ref{fig: NetworkDiagram}A) yields readily to mathematical analysis and it is straightforward to derive an implicit expression for the steady state of \bcat{} ($B^*$), along with explicit expressions for the other species. $B^*$  (given in Eqn. (\ref{eqn: bcat-stst})) exhibits qualitatively appropriate behaviour within the Wnt system. For instance, $B^*$ is increased by: increasing the Wnt stimulus; increasing the production rate of \bcat; decreasing the rate constant for formation of the destruction complex; or by decreasing the rate at which the destruction complex binds to \bcat{}. Steady state analysis also suggests that strong interaction of \bcat{} with NICD attenuates the response of \bcat{} to variation in Wnt levels (Fig.~\ref{fig: WntOnlyCrosstalk}). When the crosstalk with Notch is reduced, $B^*$ increases with the level of the Wnt signal.

Steady-state analysis of our Notch submodel (Fig.~\ref{fig: NetworkDiagram}B) is more complex. If we assume that the concentrations of NICD and the complex of NICD bound to \bcat{} are both linearly proportional to that of Notch, the model simplifies to four equations (for Notch, Hes1, proneural protein and Delta). The ensuing analysis demonstrates the existence of a steady state for Hes1 (see Eqn. (\ref{eqn: hes1-stst})), and hence for the other elements of the pathway. Linear stability analysis identifies oscillatory dynamics for Hes1, associated with the parameter $\theta_2$, which governs the balance of Wnt-mediated and Notch-mediated transcription of Hes1. In particular, our analysis suggests that the action of \bcat{} upon the Hes1 promoter serves to stabilise Notch and dampen oscillations. This agrees with an earlier mathematical model of Hes1 regulation \cite{agrawal2009computational} in which the transcriptional repression of Hes1 was varied to achieve a transition between oscillation and bistable switching.

Our studies of the decoupled Notch and Wnt systems confirm the existence of biologically realistic steady states and, in the case of Notch, regimes which yield damped oscillations. These findings are qualitatively consistent with experimentally observed behaviour for oscillations in the Notch system \cite{lewis2003autoinhibition, hirata2002oscillatory} as well as existing mathematical representations of oscillation \cite{collier1996pattern, monk2003oscillatory, agrawal2009computational}. Similarly the \bcat{} response in the Wnt system detailed here is consistent with qualitative behaviour reported in the experimental literature \cite{hernandez2012kinetic}.

\subsection*{Transcriptional regulation of Hes1 is central to dynamics of the decoupled system}
Traditional steady state analysis as described above is impeded by the complexity of the reaction network. Insights into the influence of crosstalk require alternative methods capable of analysing the full network. For example, we would like to establish whether the full model exhibits multiple steady states. Often one would like to preclude or assert particular dynamic behaviour, even if model parameters change from their estimated values, or if the functional form of terms in the model changes. \emph{Chemical reaction network theory} (CRNT) \cite{shinar2012concordant} can determine the multistationarity properties of the network without specifying either parameter values or explicit functional forms for the dynamics and without recourse to the system reductions employed in the Notch steady-state analysis described above.

CRNT identifies how coupling of the subnetworks affects the stability of the system, and achieves this via testable properties derived from the network structure. Specifically, we want to know whether a network is \emph{concordant}, a property which relates to its physical architecture in a parameter- and equation-free setting. A full definition of this property is technically involved; further details are provided in Methods and Models and a complete mathematical description, along with an illustration of the concordance property (Fig.~\ref{fig: CRNT-example}), in the SI. The \emph{Chemical Reaction Network Toolbox} \cite{ji2013crnt} provides network analysis algorithms which can assess our network -- or its constituent subnetworks -- for concordance.

Shinar and Feinberg \cite{shinar2012concordant} proved that when a network is {\em weakly monotonic} (i.e. increasing the rate of a particular reaction increases the concentration of at least one of its reactant species) and {\em concordant}, then multiple steady states are precluded, as are degenerate positive steady states. For networks satisfying additional properties ({\em weakly reversible}, {\em conservative}, continuous kinetics, and non-zero initial condition), there will be precisely one steady state, for which all species concentrations are strictly positive. In some cases (smooth kinetics in which each species has an associated degradation), concordance can assert that the unique steady state is stable, in that every real eigenvalue associated with it has negative real part. CRNT can therefore indicate the general stability properties of our reaction system by determining concordance, or lack thereof, and a broad classification for the functional forms used to describe the reaction rates of the network; it does not require a specific instantiation of the model.

Results from CRNT analysis of the decoupled and full systems of our model are given in Table \ref{tab: CRNT-results}. Inspection of our model equations (\ref{eqn: notch}) -- (\ref{eqn: axin}) confirms that our model satisfies the requirements for weakly monotonic kinetics allied with an influence specification (which describes the species which up- or down-regulate each reaction) and so the results of Shinar and Feinberg apply. Concordant subnetworks within our model are therefore monostable, whilst discordant subnetworks are multistable \cite{shinar2012concordant}.

\paragraph*{Notch subnetwork identified as discordant and translates to full system.}
\begin{table}[!ht]
\begin{tabular}{|l|c|c|c|}
\hline
{\bf Model} & {\bf Active Coupling Points} & {\bf Wnt ON} & {\bf Wnt OFF}\\ 
\hline
\multirow{2}{*}{\emph{Notch Only}} & Step \mycirc{4} only & \tick & \cross \\
& Step \mycirc{4} \& Step \mycirc{14} & \tick & \tick \\
\hline
\emph{Wnt Only} & N/A & \tick & \tick \\
\hline
\multirow{2}{*}{\emph{Full System}} & Step \mycirc{4} only & \cross & \cross \\
& Step \mycirc{4} \& Step \mycirc{14} & \tick & \tick \\
\hline
\end{tabular}
\caption{{\bf Chemical reaction network results for decoupled and full system.} Concordance results for decoupled and whole networks in a homogeneous system, analysed using the CRN Toolbox\cite{ji2013crnt}. A tick indicates a concordant network (monostable) and a cross, a discordant network (multistable). The second column indicates which \bcat{} crosstalk points were included in each network. \bcat{} is not consumed in Step 14 and so the comparison of the two coupling points does not apply to the Wnt-only system.}
\label{tab: CRNT-results}
\end{table}

Comparisons between the decoupled and full systems yield three valuable insights into the origins of the behaviour of the coupled network:
\begin{enumerate}
\item{{\bf{\bcat{} acts as a stabilising force:}}} Direct action of \bcat{} upon the Hes1 promoter (networks involving Step \mycirc{14}) always yields an injective, and hence monostable, network. \bcat{} serves to dampen the multistability associated with the Notch system. Conversely, all discordant networks in this analysis arise when Hes1 transcription is solely Notch-mediated.
\item{{\bf{Discordance arising from Wnt response:}}} In the presence of a Wnt stimulus, monostable Notch and Wnt subnetworks can combine to produce a multistable network. This occurs when Hes1 transcription is solely Notch-mediated. Harrington \etal{} \cite{harrington2013cellular} demonstrate that in such cases, the rate of shuttling of crosstalk species (in our case, \bcat) between the subnetworks determines the region of parameter space in which multistability occurs.
\item{{\bf{Multistability in Notch influences the full system:}}} Sub-networks can transmit their discordance to the full network, as shown by Shiu \cite{shiu2008smallest}. Some of the discordance in the full system (Step \mycirc{4} only, Wnt off) can be attributed to the equivalent discordance in the Notch system.
\end{enumerate}
The results in Table \ref{tab: CRNT-results} show that the stability properties of the full system depend not just on the presence or absence of a Wnt stimulus, but also on the mechanisms governing transcription of Hes1, Steps \mycirc{4} and \mycirc{14} of the model. In particular, the relative contribution of Notch- and Wnt-mediated transcription of Hes1 determines whether the full system is monostable or multistable. Biologically this corresponds to Notch favouring heterogeneity and flexibility for cell fate selection, whereas Wnt influences the system towards a single steady state. 

In light of these findings, the subsequent two-cell studies focus upon the relative contribution of Notch- and Wnt-mediated regulation of Hes1 and the strength of the Wnt stimulus $W$, with a view to understanding how activity around the Hes1 crosstalk hub delivers the coordination of fate selection seen in tissues of the intestinal epithelium.

\subsection*{Implications of cross-talk for heterogeneous states}
The next application focuses upon \emph{in silico} studies of a healthy cell pair using the complete network (both Notch and Wnt pathways, and their crosstalk) shown in Fig.~\ref{fig: NetworkDiagram}A-C, in which the governing parameters for Hes1 transcription can be manipulated to deliver either Wnt-dominant or Notch-dominant control of the Hes1 promoter. Since cell populations in tissues are naturally heterogeneous, this focuses on {\em two coupled heterogeneous cells}, each running an embedded system of the twelve model ODEs, using the parametrisation described in Tables \ref{tab: dimensional-decay-rates} -- \ref{tab: general-rate-params}. All cells start from the standard Wnt conditions listed in Table \ref{tab: initial-conditions}; the first cell of the pair adopts conditions of $0.5nM$ for all its Notch components, whilst the Notch entities of the second cell start from $0.51nM$. This difference permits the emergence of heterogeneous states.

\begin{figure}[htbp]
\centering
\includegraphics[width=\textwidth]{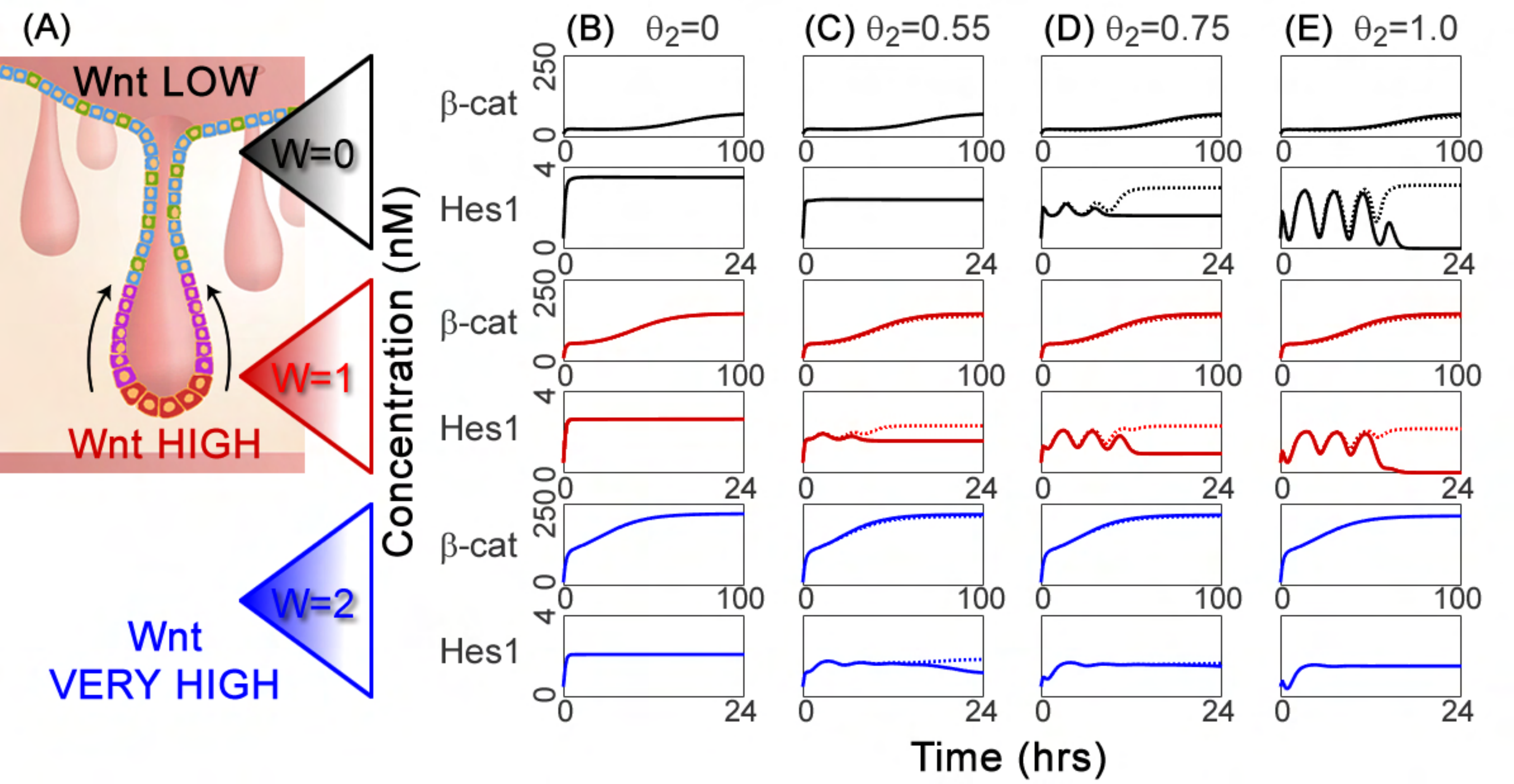}
\caption[]{{\bf Plots showing how variation in $\theta_2$, the proportion of Notch-mediated control of the Hes1 promoter, affects the system dynamics.} (A) Cross-section schematic of a crypt from the large intestine; image adapted from Reizel \etal{} \cite{reizel2011colon}, originally published by PLoS and provided under a Creative Commons Attribution Licence, \emph{CC-BY-2.5}. (B--E) Influence of $\theta_2$ upon the Hes1 steady state at $W=0.0$ (black), $W=1.0$ (red) and $W=2.0$ (blue); $\theta_2 \in [0,1]$ represents the proportion of Notch-mediated transcription of Hes1. Timecourses show Hes1 and \bcat{} expression for healthy cell pairs, for (B) $\theta_2 = 0.00$ (i.e. entirely Wnt-mediated), (C) $\theta_2 = 0.55$, (D) $\theta_2 = 0.75$, (E) $\theta_2 = 1.00$ (i.e. entirely Notch-mediated). The timecourse for the first cell of each pair is indicated by a solid line; the second, by a dotted line. Where only one timecourse is apparent, the cell pair is synchronised. Standard initial conditions and parameters are used, stated in Tables \ref{tab: dimensional-decay-rates} -- \ref{tab: initial-conditions}; the ODE model comprises equations (\ref{eqn: notch}) -- (\ref{eqn: axin}).}
\label{fig: theta2}
\end{figure}

Simulation results from these healthy cell pairs are displayed in Fig.~\ref{fig: theta2}B--E. In our model, the parameter $\theta_2$ directly reflects the proportion of Notch-mediated control of the Hes1 promoter (Eqn. (\ref{eqn: hes1})). Fig.~\ref{fig: theta2} demonstrates the change in expression of \bcat{} and \hes{} as this proportion is varied, ranging from  $\theta_2=0.0$ (regulation wholly Wnt-mediated), to $\theta_2=1.0$ (regulation entirely Notch-mediated). Simulations are presented for Wnt stimulus absent (e.g. crypt orifice, $W=0$), present (e.g. crypt base, $W=1$) or excess (e.g. hyperstimulated conditions, $W=2$), aiming to mimic broad differences in the Wnt stimulus observed on ascending the intestinal crypt.

\paragraph*{Relative Contribution of Notch- and Wnt-Mediated Action on the Hes1 Promoter is Key to Shaping Dynamics}
\label{sec: theta2}
The relative strength of the two Hes1 regulation mechanisms strongly influences the dynamics of the Notch system.  Oscillations occur where Hes1 transcription is substantially Notch-mediated (Fig.~\ref{fig: theta2}D, E). When Wnt-mediated control dominates (Fig.~\ref{fig: theta2}B) or the two routes are close to parity (Fig.~\ref{fig: theta2}C), oscillations are more strongly damped and both cells settle rapidly on a constant steady state, characterised by reduced levels of Hes1.

It is noteworthy that the steady state of \bcat{} appears unaffected by interaction with the Notch pathway, although some simulations at low-Wnt, high-$\theta_2$ develop small-amplitude oscillations about the steady state in response to the oscillations in the Notch system. 

These results are consistent with the idea of Notch signalling maintaining flexibility in cell fate decisions \cite{kageyama2008dynamic}. Cells could regulate this by preserving a Notch-mediated monopoly of the Hes1 promoter, allowing Notch oscillations to persist until a \bcat{} intervention at a specified time induces bistability and the commencement of fate selection. 

\paragraph{Higher Hes1 Levels Correlate With Reduced \bcat{}}
Where slight differences emerge between the two cells' \bcat{} levels (Fig.~\ref{fig: theta2}C--E for $W = 0, 1$), higher Hes1 levels correlate with marginally lower \bcat{} expression. This may be due to \bcat{} being diverted for use in binding with NICD (Step \mycirc{3} of Fig.~\ref{fig: NetworkDiagram}B) in the Notch-mediated route for Hes1 regulation. The stronger the Wnt stimulus, the more \bcat{} is expressed, but the qualitative form of the \bcat{} timecourse remains the same.

\paragraph*{Notch-Wnt Interaction Supports Spatial Coordination in the Crypt}
Oscillation of the Notch system is also regulated by the extracellular Wnt stimulus. Hyperstimulation ($W=2$) dampens oscillations and promotes homogeneity in Hes1 expression. This is due to the Wnt-dependent downregulation of Hes1 included in our model, representing Dsh-mediated activity at the Hes1 promoter (described in the Notch Pathway Submodel of the SI). Furthermore, the transition from the Wnt-on ($W=1$) state at the crypt base to the Wnt-off ($W=0$) state at the orifice sees an increase in the maximum expression of Hes1 and an increase in the difference in Hes1 expression between each cell of the pair. These results are consistent with the emergence of cell heterogeneity in the intestinal epithelium as cells traffic up the crypt axis from a high to a low Wnt stimulus.

\paragraph*{} Notch-mediated control of the Hes1 promoter alone does not suffice to drive cell fate selection in the intestinal crypt. Rather, for Hes1 oscillations and heterogeneous cell fates to occur, we require coordination between Notch-mediated control of the promoter and a drop in the Wnt stimulus. Key to this coordination is the Dishevelled-mediated downregulation of Hes1 in the face of a high extracellular Wnt stimulus \cite{collu2012dishevelled}, which effectively creates a Wnt-dependent `sweet spot' for Hes1 oscillations. This mechanism could offer an important role in linking the spatial gradients of extracellular Wnt to the coordination of emerging cellular heterogeneity within the intestinal crypt epithelium.

\subsection*{Cell mutations are affected by Notch-Wnt crosstalk}
All results presented thus far have focused on healthy cells. We now update the model to approximate the altered biochemistry of a `mutant' or modified phenotype, with particular reference to the cells of the intestinal epithelium. 

The first modified phenotype is an APC mutant, commonly implicated in colorectal cancer \cite{kinzler1996lessons}, in which the function of the \bcat{} destruction complex is either partially or wholly impaired and therefore has a reduced binding affinity for \bcat. APC is coded for genetically by two alleles, each of which is either healthy or mutated; the majority of CRC tumours display inactivation of both alleles \cite{nagase1993mutations}. APC is represented implicitly in our model, via the formation of the destruction complex $C$. We emulate APC mutation via a multiplier $\rho_{APC} \in [0,1]$, which scales the rate of formation of the destruction complex in Eqns. (\ref{eqn: gsk}) and (\ref{eqn: dest-complex}) for $G$ and $C$ respectively. Healthy cells possess two normal copies of the APC allele and therefore have $\rho_{APC} = 1.0$. A single-hit APC mutation, in which one allele is mutated, takes $\rho_{APC} = 0.5$, whilst a two-hit mutation sets $\rho_{APC} = 0.0$.

The second modified phenotype confers a hyperstimulated Wnt state and is motivated by suggestions in the literature that differential response to Wnt stimuli, rather than differential exposure, is a major influence upon cell proliferation \cite{sato2009single}. The hyperstimulated phenotype is implemented by changing the value of the Wnt stimulus $W$ for the affected cell from the reference state $W=1.0$, to $W=2.0$. This study aims to explore how variability in the response to local Wnt stimulus across a cell population might impact upon the expression of cell fate determinants such as Hes1.

\begin{figure}[p]
\centering
\includegraphics[width=\textwidth]{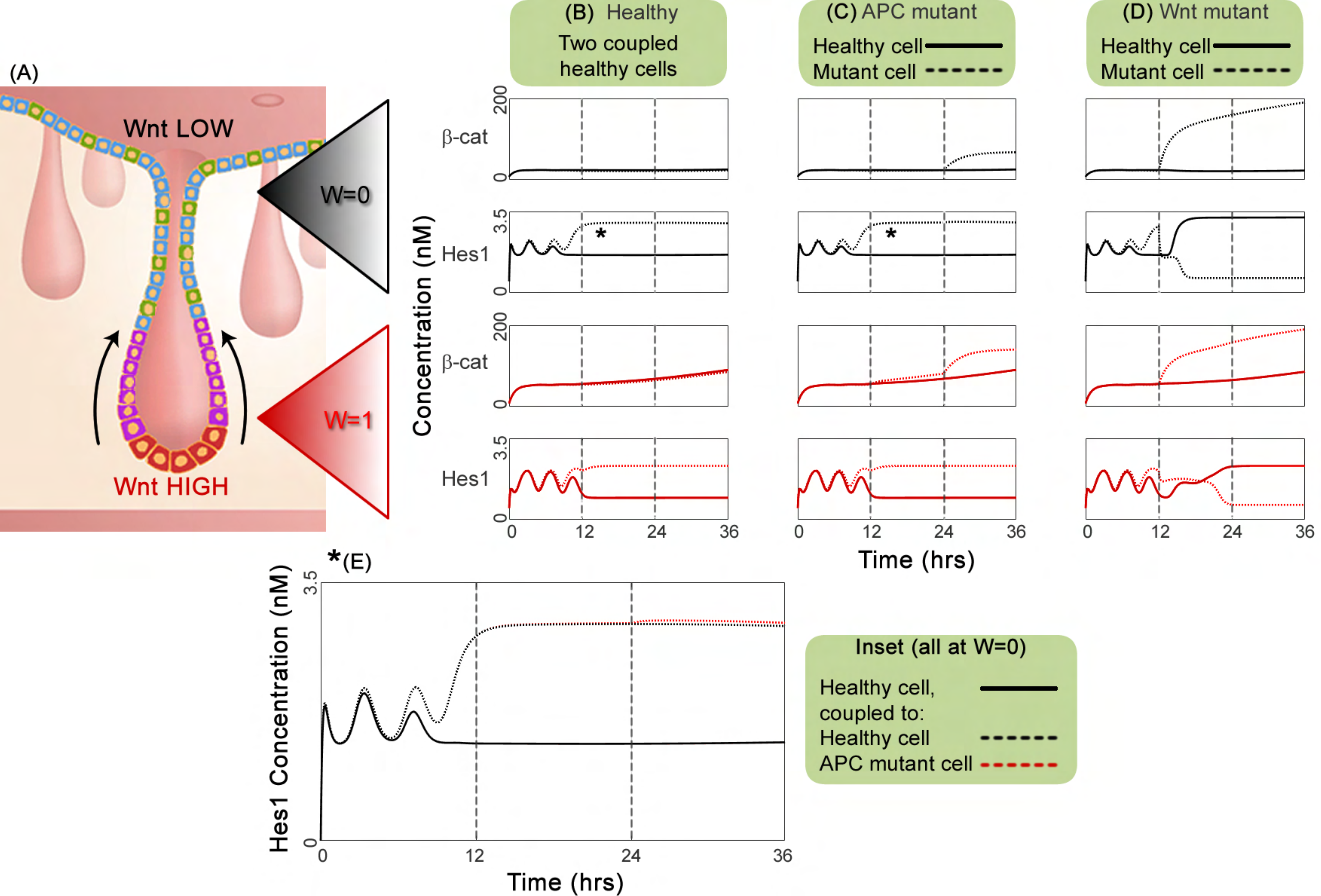}
\caption{{\bf Dynamic timecourses showing cell pair responses to mutation in one cell} (A) Cross-section schematic of a crypt from the large intestine; image adapted from Reizel \etal{} \cite{reizel2011colon}, originally published by PLoS and provided under a Creative Commons Attribution Licence, \emph{CC-BY-2.5}. (B--E) Timecourses for cell pairs started from homogeneous initial conditions; all cells are healthy at start of simulation. (B) Healthy cell pair; (C) APC mutants (dashed line) acquire their first APC knockout ($\rho_{APC} = 0.5$) at $t=12h$ and the second ($\rho_{APC} = 0.0$) at $t=24h$; (D) hyperstimulated Wnt mutants (dashed line) transform to a $W=2.0$ state at $t=12h$. Except for Wnt mutants, all plots in panels (B)--(D) indicate simulations with (red) $W=1.0$ and (black) $W=0.0$. Inset panel (E) compares the Hes1 expression in the healthy and APC mutant scenarios for $W=0$, indicated by the asterisks. In this case, the healthy scenario is shown in black and the APC mutant in red. Standard initial conditions and parameters are used, stated in Tables \ref{tab: dimensional-decay-rates} -- \ref{tab: initial-conditions}; the ODE model comprises equations (\ref{eqn: notch}) -- (\ref{eqn: axin}), with APC modifications to Eqns. (\ref{eqn: gsk}) and (\ref{eqn: dest-complex}) for study (C).}
\label{fig: mutant-hits}
\end{figure}

Results for two-cell simulations of these modified phenotypes are shown in Fig.~\ref{fig: mutant-hits}B--E. In all cases except the hyperstimulated Wnt mutant, we fix $W=0.0$ (Wnt stimulus off, black plots) or $W=1.0$ (Wnt stimulus on, red plots), to simulate conditions near the crypt orifice or base respectively. Cell pairs are healthy at $t=0h$ and are initialised as for the previous two-cell simulations. In the APC mutant study, one cell acquires a single APC mutation at $t=12h$ and a second hit at $t=24h$. For the Wnt mutant study, one cell mutates to a hyperstimulated state at $t=12h$.

\paragraph*{Role of Notch-Wnt Crosstalk in Counteracting Effects of APC Mutation}
Fig.~\ref{fig: mutant-hits}C reveals that the first APC hit breaks the symmetry of the cells' \bcat{} dynamics and these differences are magnified once the second hit appears. The acquisition of APC mutation(s) reduces the rate of destruction of \bcat, thereby increasing its expression within the cell; this effect is more pronounced for the high-Wnt conditions associated with lower regions of the crypt. Two mutations are required for a substantial departure from the healthy state. 

Some small changes in Hes1 expression occur, as shown in the inset of Fig.~\ref{fig: mutant-hits}E, although these only become apparent after the second APC hit. Impairment of the destruction complex enables the mutant phenotype to express higher levels of Hes1, associated with prolonged mitotic activity, and reduced levels of Hath1 (data not shown), which promotes cell-cycle exit and is associated with further cell fate specification. Elevated levels of \bcat{} in the APC mutant upregulate Hes1, via Notch- and Wnt-mediated routes. That this is more pronounced for $W=0$ may indicate a role for the Dsh-mediated downregulation of Hes1 in counteracting the effects of mutation near the crypt base, when the local Wnt stimulus is high.

\paragraph*{Wnt Hyperstimulation Induces Cell Fate Switching}
The initial timecourse for Hes1 in Fig.~\ref{fig: mutant-hits}D oscillates and demonstrates mild divergence by $t=12h$, in both the upper ($W=0$) and lower ($W=1$) crypt. At this point, the second cell has the higher Hes1 expression. Following hyperstimulation of the second cell, there is an abrupt change in the pattern of Hes1 expression; Hes1 levels in the second cell fall sharply, while those in the healthy cell increase. Hes1 oscillations cease and the cells evolve to a constant steady state, with a high-Hes1 healthy cell and a low-Hes1 mutant. Hyperstimulation of the Wnt pathway enables the mutant cell to invert the cell fate decision of its neighbour, forcing it from a primary to a secondary fate. Although this behaviour is evident in both the Wnt-on and Wnt-off scenarios, the transitions post-mutation are more pronounced and occur over shorter times when $W=0$. This reversal of roles affects all variables in the Notch submodel (data not shown). The Wnt hyperstimulated mutants display substantially elevated \bcat{} expression compared to the healthy case (Fig.~\ref{fig: mutant-hits}B): by $t = 24h$, a $2.2$-fold increase at $W=1$, and an $8.3$-fold increase at $W=0$.

The observed reversal of Hes1 expression patterns in the cell pair may have consequences for fate selection in a multicellular environment. Given that Hes1 expression is associated with maintaining a proliferative phenotype, the ability of a mutant cell to invert cell fate decisions could stimulate surrounding cells to continue in a mitotically active state for longer. The elevated \bcat{} expression of hyperstimulated mutants would also help to maintain active cycling. Consequently, mutation events might generate mitotically active clusters which are only partly composed of aberrant cells, as for example in the Cancer Stem Cell Hypothesis \cite{salama2009colorectal, papailiou2010stem}.

\section*{Discussion}
Our mathematical model for Notch-Wnt crosstalk captures the main qualitative features of each pathway, such as the Notch pathway's capacity for damped oscillations and the Wnt pathway's regulation of \bcat{} expression by the extracellular Wnt concentration, and provides good agreement with the available experimental data \cite{hirata2002oscillatory, hernandez2012kinetic}. Computational exploration of our model, whether through the powerful abstractions of CRNT or the \emph{in silico} simulation of cell pairs using a parameter- and dynamic-specific instantiation, has demonstrated how a nuanced balance of Notch- and Wnt-mediated regulation of the Hes1 promoter shapes the timing and outcome of cell fate selection in the intestinal crypt epithelium. The following principal findings have emerged through analysis and simulation of either homogeneous cell populations or heterogeneous cell pairs:
\begin{itemize}
 \item \textbf{Wnt stabilises Notch:} Direct action of \bcat{} on the Hes1 promoter confers a single steady state on a homogeneous Notch-Wnt network, dampening oscillatory dynamics in the Notch pathway. \bcat{} crosstalk stabilises the output of the Notch pathway and reduces the flexibility of fate decision which would otherwise be conferred by oscillations in Hes1.
 \item \textbf{Relative contribution of Notch- and Wnt-mediated control of the Hes1 promoter shapes Notch dynamics:} The presence or absence of oscillations is associated with control of the Hes1 promoter. Notch-mediated regulation of Hes1 transcription promotes oscillations, while Wnt-mediated regulation via direct binding of \bcat{} to the Hes1 promoter dampens oscillations and induces the cell to settle on a constant steady state. Furthermore, Wnt-induced downregulation of Hes1 via the interaction of Dishevelled with the Hes1 promoter may serve to prevent oscillations in regions where the Wnt stimulus is too high.
 \item \textbf{Role for Notch-Wnt crosstalk in counteracting the effects of APC mutation:} APC mutation impairs the action of the \bcat{} destruction complex, increasing the expression of \bcat{} and hence Hes1, via Notch- and Wnt-mediated transcription routes. Effects on Hes1 are marginally more pronounced in the low-Wnt conditions of the upper crypt. This may indicate a role for the Wnt-related downregulation of Hes1 via Dishevelled, in buffering the effects of mutations downstream of Dsh in the Wnt pathway, in the high-Wnt conditions of the lower crypt.
 \item \textbf{Wnt hyperstimulation can determine the fate of neighbouring cells:} Our simulations of cell pairs suggest that a Wnt-hyperstimulated cell may drive neighbouring cells to adopt a secondary, low-Delta fate. 
\end{itemize}
We now discuss the biological implications of our findings. For example, cells might regulate the relative contribution of Notch- and Wnt-mediated transcription routes in order to coordinate the timing of cell fate selection. As cells migrate up the crypt, Wnt levels and hence \bcat{} expression fall, enabling a shift towards Notch-mediated control of the Hes1 promoter and favouring the emergence of distinct cell fates. 

Other features emerging from our model include the fate reversal seen in healthy cells neighbouring a Wnt-hyperstimulated phenotype, which could have some relevance to the cancer stem cell hypothesis, such that a hyperstimulated cell could maintain neighbouring healthy cells in a mitotically active state.

The success of chemical reaction network theory in identifying the emergence of full-system multistability from two monostable subnetworks highlights the importance of including pathway crosstalk in our mathematical models, if the richness of the underlying biochemistry is to be captured. Very different dynamics are obtained when the crosstalk between the Notch and Wnt pathways is accounted for, and our model has demonstrated that the Wnt pathway in particular has substantial capacity to influence the outcomes of Notch signalling. Crosstalk between the pathways should therefore be included in any future mathematical models where both proliferation and cell fate specification are being investigated.

Simulations of cell pairs have yielded useful insights into healthy and aberrant scenarios. Future work would need to extend these studies to larger cell populations, preferably within a geometrically realistic crypt setting, to explore how the mutations described manifest at tissue level and over longer timescales. Full-crypt simulations (capturing three-dimensional populations of crypt cells, as for example in \cite{leeuwen2009integrative, mirams2012theoretical}) to extend the Wnt hyperstimulation study might also examine whether the cancer stem cell hypothesis emerges from our model in larger populations. Given our focus upon the Hes1 promoter, it may prove beneficial to refine this area of the model to incorporate Hes1 mRNA and dimerisation as in other mathematical models \cite{momiji2008dissecting, momiji2009oscillatory, agrawal2009computational, kiparissides2011modelling}, to enable closer matching of the oscillatory readings with experimental data or, alternatively, refinement of the Wnt submodel to include explicit representation of Dishevelled \cite{lee2003roles, benary2015mathematical}.

\section*{Methods and Models}

\subsection*{ODE Solvers}
Numerical solution of ODEs, involved in the steady-state analysis and all cell pair simulations, is perfomed in \matlab{}, using the software's own suite of solvers \cite{shampine1997matlab}. Owing to the stiffness of the ODE model, we employ the solver \texttt{ode15s}: this is a multistep, variable order solver, and employs an algorithm based upon the numerical differentiation formulas \cite{shampine1997matlab}.

\subsection*{Chemical reaction network theory analysis}
All CRNT analyses are performed using the \emph{Chemical Reaction Network Toolbox}, a computational package for analysing the stability properties of chemical reaction networks, indicating whether a given network is capable of multiple stable states, or only one \cite{ji2013crnt}. The toolbox requires the user to specify the details of each reaction in the network of interest. Each inclusion takes the form \ce{A + B ->[\ce{E}] 2C + D}, where A and B are the reactants, C and D are the products, and E is either an inhibitor or a promoter of the reaction.

We supply the toolbox with the species (Table \ref{tab: entities}) and network connectivity of the Notch subnetwork, Wnt subnetwork and fully coupled network (Fig.~\ref{fig: NetworkDiagram}), along with the \emph{influence specification}, the species which up- or down-regulate each interaction.  We also provide two versions of the Notch subnetwork and whole network. The first version contains both Wnt- and Notch-mediated regulation of Hes1 (Steps \mycirc{14} and \mycirc{4} of Fig.~\ref{fig: NetworkDiagram} respectively), while the second involves only Notch-mediated control. The toolbox also requires stoichiometry information; for instance, specifying 2C rather than just C as a product in the above example. However, the kinetics of each reaction (i.e. the functional forms of the reaction rates) and parameter values do not need to be specified: the toolbox would not require a specific functional form for the dependence of the reaction rate upon E in the above example. The toolbox performs a sign-checking operation on two quantities derived from the network's stoichiometry \cite{ji2013crnt} and classifies the network according to its ability to exhibit multistationarity if allied with particular types of kinetics. 

It is not possible to analyse states of the heterogeneous system, owing to the excessively large computation time for a system of this complexity; consequently, all our CRNT results relate to a homogeneous Notch pathway (in which we assume $D = \bar{D}$ in Eqn. (\ref{eqn: notch})). Crosstalk species in the decoupled systems are represented as full reactants with their own inflows and outflows. 

\subsection*{Parametrisation}
Computational implementation of our model requires us to determine appropriate parameter values. Some of our $41$ parameters, such as decay rates, are readily amenable to experimental measurement, whilst others, such as binding rates, are not. Experimental estimates are not available for $21$ of our model parameters at the present time. Consequently:
\begin{enumerate}
 \item Estimates derived from human cell lines, in particular intestinal epithelial lines, have been used wherever possible;
 \item Where data from human cell lines is absent, values from mammalian lines have been employed where possible;
 \item Otherwise, non-mammalian readings or values from published mathematical models have been used as initial estimates for parameter fitting studies.
\end{enumerate}
Our primary focus is on the qualitative features of the model within biologically realistic regimes. Nonetheless, it might be hoped that qualitative predictions from our model could stimulate future experimental estimation of its parameters within a single, human intestinal cell line. 

\begin{figure}[thbp]
\centering
\includegraphics[width=\textwidth]{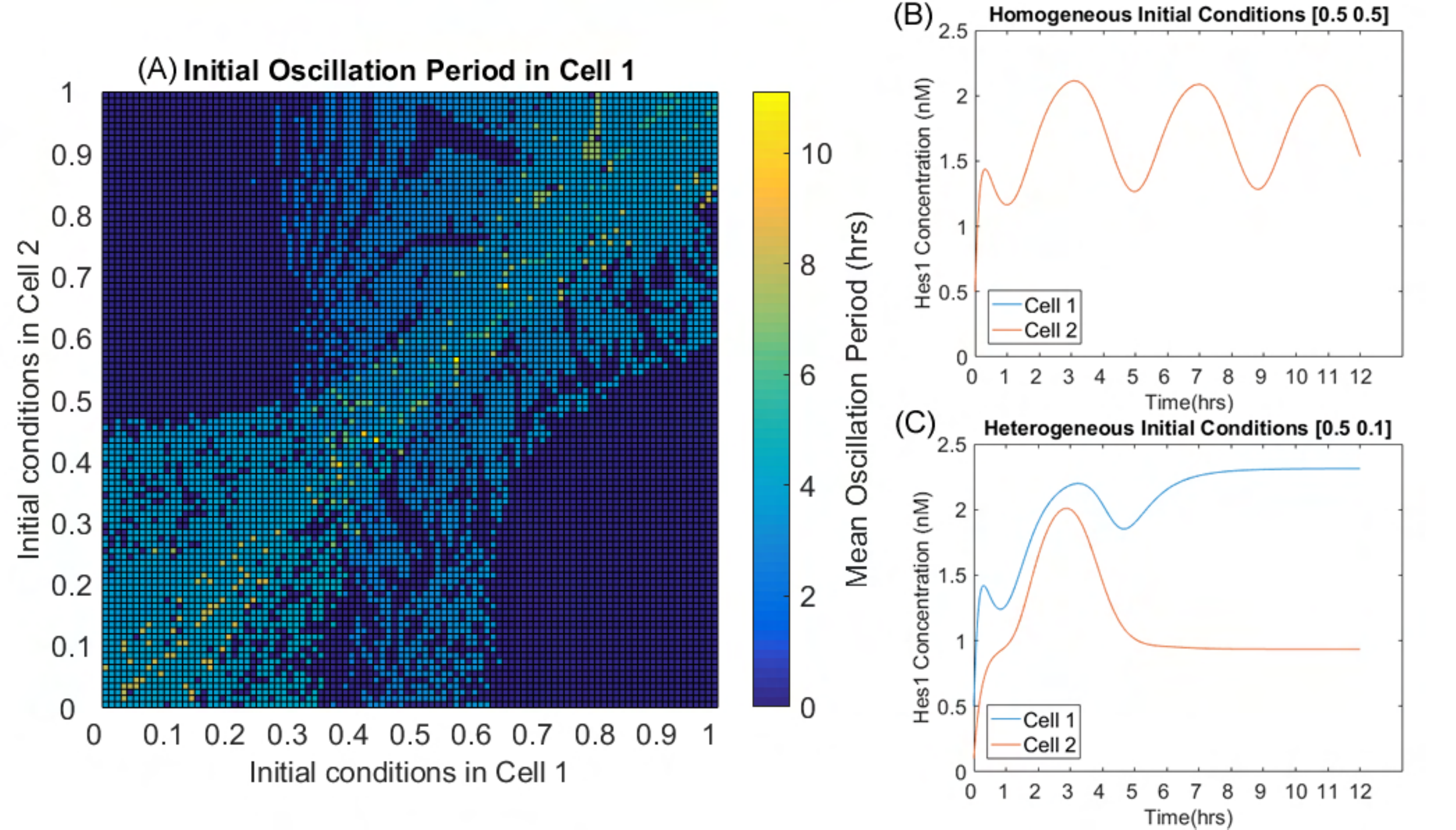}
\caption[]{{\bf{Notch parametrisation outcomes}} (A) Response of oscillation period in cell $1$ to variations in initial conditions in a two-cell system running the decoupled, dimensional Notch model, Eqns.(\ref{eqn: notch}) -- (\ref{eqn: delta}). Statements of initial conditions of the form $[x,y]$ indicate that all seven Notch species in cells $1$ and $2$ are initialised to $x$ and $y$ respectively. Owing to symmetry considerations, only the oscillations in Cell 1 were measured; results for Cell 2 correspond to a reflection of this surface in the line $y=x$. An amplitude filter was applied during generation of the plot, to disregard any small-amplitude oscillations ($< 0.001$) arising from the computational solution process, rather than true oscillations of the ODE model. (B) Diagonal entries of (A) yield homogeneous evolution with damped oscillations, as in this timecourse of a cell pair from initial conditions ($0.5, 0.5$). (C) Off-diagonal entries of (A) show heterogeneous evolution, as in this timecourse of a cell pair from initial conditions ($0.1, 0.5$).}
\label{fig: hirata-matching}
\end{figure}

\begin{figure}[h]
\centering
\includegraphics[width=\textwidth]{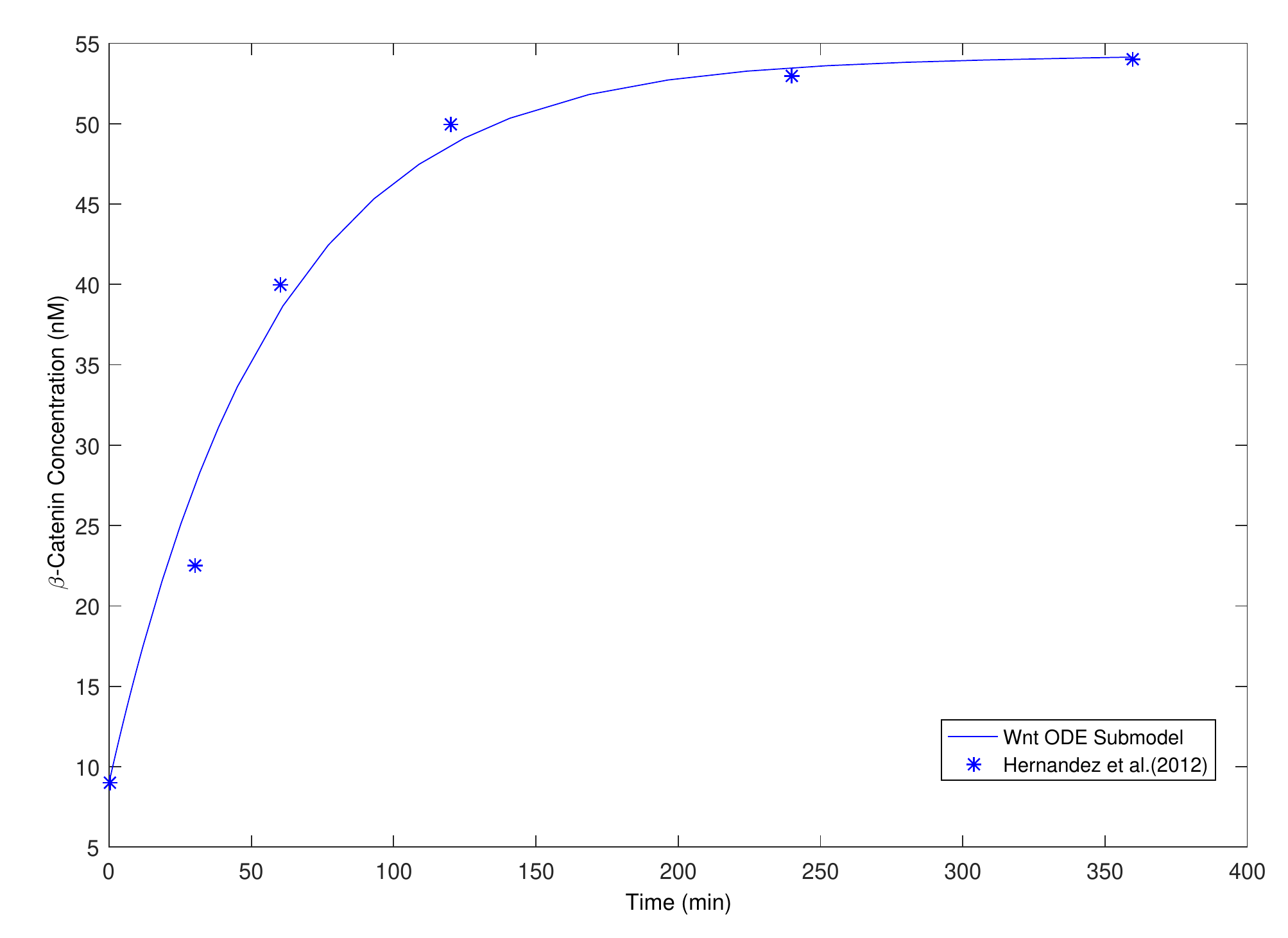}
\caption[]{{\bf{Wnt parametrisation outcome}} Timecourse for \bcat{} evolution in the dimensional, decoupled Wnt submodel, Eqns. (\ref{eqn: gsk}) -- (\ref{eqn: axin}) (line graph), compared against the experimental readings of Hern\'andez \etal{} \cite{hernandez2012kinetic} (point data). The timecourse for the Wnt submodel uses the parameter values listed in Tables \ref{tab: dimensional-decay-rates} -- \ref{tab: general-rate-params}, and the initial conditions of Table \ref{tab: initial-conditions}.}
\label{fig: hernandez-matching}
\end{figure}

All parameter fitting uses the decoupled Notch and Wnt systems; fitted parameters are listed in full in the SI, indicated by a `\emph{PF}' in Tables \ref{tab: dimensional-decay-rates} -- \ref{tab: general-rate-params}. In each case, an initial set of parameter values is formed, applying criteria $1 - 3$ above. Using the Systems Biology Toolbox, an add-on kit for \matlab \cite{schmidt2006systems, shampine1997matlab}, a sensitivity analysis is performed on this set (Fig.~\ref{fig: Sensitivity}) to determine the order in which parameters are to be sequentially varied. As a given parameter is varied, the resulting model output is measured against the target data (Hes1 oscillation period in the Notch system, \bcat{} steady state in the Wnt system) and the new value selected which delivers the closest match to the data mentioned below. Parameters are modified sequentially on a loop until the model output lies within a given tolerance of the target data. Although this approach to obtaining a parameter set is unlikely to yield the global optimum, it nonetheless provides a parametrisation which is fully grounded in the literature, biologically plausible, and results in biologically plausible behaviour, as evidenced by the matching to experimental data (shown in Figs. \ref{fig: hirata-matching} and \ref{fig: hernandez-matching}). Full details of tolerances, initial estimates and the final parameter values are stated in Tables \ref{tab: dimensional-decay-rates} -- \ref{tab: initial-conditions}.

Parametrisation of the Notch submodel defined by Eqns. (\ref{eqn: notch}) -- (\ref{eqn: hath1}) targets the two hour period of Hes1 oscillations observed by Hirata \etal{} \cite{hirata2002oscillatory} in murine myoblast cells. Initial parameter estimates were determined by solving for the homogeneous state of a two-cell system. Data from the model of Shepherd \cite{shepherd2010thesis} were used to locate an oscillatory regime of our Notch model, yielding suitable initial estimates for missing parameters, while initial conditions for model variables were chosen to be of the same order as in the model of Agrawal \etal \cite{agrawal2009computational}. On completion of the fitting procedure, the parameter set was tested in a two-cell, heterogeneous system and the oscillation period measured over a range of starting conditions. Where oscillations occur, the period generally lies within the $2 - 4$ hour range (Fig.~\ref{fig: hirata-matching}). This offers a reasonable match to the data of Hirata \etal{} but tends to overestimate the oscillation period, as expected for a non-delay model of this kind \cite{monk2003oscillatory, momiji2008dissecting}.

Parametrisation of the Wnt submodel described by Eqns. (\ref{eqn: gsk}) -- (\ref{eqn: axin}) uses the data of Hern\'andez \etal{} \cite{hernandez2012kinetic}, which supplies a \bcat{} timecourse for the human colon carcinoma cell line, $RKO$. Consequently, all Wnt stimuli in our model are nondimensionalised against a reference value of \mbox{$100$ ng/ml}. The unstimulated state, $W=0$ in our model, equates to $0$ ng/ml; the reference value represents $W=1$. All other values scale linearly with this, with values $W>1$ representing a hyperstimulated state. Steady-state data from Hern\'andez \etal{} \cite{hernandez2012kinetic} are used to generate a pair of simultaneous equations (Eqns. (\ref{eqn: hernandez-plateau}), (\ref{eqn: hernandez-slope})) which supply estimates for two unknown parameters, $\alpha_3, \alpha_4$. Thereafter we fit the three-hour time course for \bcat. The mean squared error (MSE) of the model's performance, $\hat{X}$, against the experimental data, $X$, is calculated in each case:
\begin{eqnarray}
\label{eqn: MSE}
 MSE = \frac{1}{n} \sum_1^n (\hat{X} - X)^2 \,, \nonumber
\end{eqnarray}
where $n=6$, the total number of observations. Estimates for the initial concentrations of reactant variables are drawn either from the experimental work of Tan \etal{} \cite{tan2012wnt} which uses three human cell lines, or from the mathematical model of Lee \etal{} \cite{lee2003roles}, based on \emph{Xenopus} oocytes.
The resulting \bcat{} evolution of our Wnt submodel, shown in Fig.~\ref{fig: hernandez-matching}, provides a close fit to the data of Hern\'andez \etal{} \cite{hernandez2012kinetic}.

In general our parameter values are of the same order of magnitude as those of other Notch and Wnt models in the literature \cite{lee2003roles, mirams2010multiple, monk2003oscillatory, agrawal2009computational}. Key differences occur where our estimates benefit from more recent experimental data \cite{hernandez2012kinetic}, for example in adopting the \bcat{} decay rate $\mu_B = 0.00636 \times 10^{-3}$ min$^{-1}$ rather than the $\mu_B = 2.57 \times 10^{-4}$ used elsewhere \cite{lloyd2013toward}.

Comparisons of our decoupled models against experimental data (see Figs. \ref{fig: hirata-matching}, \ref{fig: hernandez-matching}) indicate that they capture the qualitative features of the Notch and Wnt systems and do so within a biologically sound regime. This enables us to apply the full, coupled model to problems of biological and biochemical interest.

\section*{Acknowledgments}
Many thanks to Anne Shiu for useful discussion and comments on the chemical reaction network theory aspects of the text, and to Alastair Kay for technical assistance.

\setcounter{equation}{0}
\renewcommand\theequation{S.\arabic{equation}}
\renewcommand\thefigure{S.\arabic{figure}}
\setcounter{table}{0}
\renewcommand\thetable{S.\arabic{table}}

\section*{Supporting Information}
\label{SI_Text}
The following supplementary information details our mathematical model for Notch-Wnt interaction, through a statement of the twelve differential equations alongside the associated experimental evidence. All numbered steps coincide with the numbering of Fig.~\ref{fig: NetworkDiagram}. We also introduce the parameters associated with the dimensional version of our model, and outline the steady state analysis of the decoupled systems. The ODEs (\ref{eqn: notch}) -- (\ref{eqn: axin}) comprise our model of Notch-Wnt interaction for use in the cell pair simulations.

\subsection*{Model Development}
\label{sec: equations}
We now describe the ODEs which we use to model the Notch-Wnt interaction network. With a slight abuse of notation, we denote by $X$ ``the concentration of reactant $X$'', rather than $[X]$.

The system comprises twelve ODEs: six for the Notch pathway, four for the Wnt pathway, and two for intermediate complexes which either mediate interactions between the two pathways ($I_1$) or which respond to the strength of extracellular Wnt signalling to regulate the level of active \bcat{} in the cell ($I_2$). The dependent variables, abbreviating letters and associated parameters can be found in Tables \ref{tab: entities} and \ref{tab: parameters} respectively.

We represent the extracellular Wnt stimulus by a time-dependent, nondimensional quantity, $W(t)$. Other nondimensional variables and parameters are indicated by the dash notation $'$, except where specified otherwise.

\begin{table*}[t]
\centering
\renewcommand{\arraystretch}{1.5}
\begin{tabular}{|c|c|c|c|}
\hline
\textbf{Variable} & \textbf{Description} & \textbf{Units} &\textbf{Scaling} \\
\hline
$t$ & Time & min & $\tau = \mu_N t$\\
\hline
$N$	& Membrane-bound Notch receptor & nM & $N = \frac{\theta_1}{\mu_N}N'$\\
$F$ &  Notch Intracellular Domain (NICD) & nM & $F = \frac{\theta_1}{\mu_F}F'$\\
$H_1$ &  Hes1 & nM & $H_1 = \frac{\xi_2}{\mu_{H_1}}H_1'$\\
$P$ & Ngn3 & nM & $P = \frac{\xi_3}{\mu_P}P'$ \\
$D$	& Delta ligand & nM & $D = \frac{\theta_4}{\mu_D}D'$ \\
$H_2$ &  Hath1 & nM & $H_2 = \frac{\xi_5}{\mu_{H_2}}H_2'$ \\
\hline
$G$ & \gsk & nM & $G = \frac{\theta_1}{\mu_G}G'$ \\
$C$ & Destruction complex & nM & $C = \frac{\theta_1}{\mu_C}C'$ \\
$B$ & Active \bcat & nM & $B = \frac{\theta_1}{\mu_N}B'$ \\
$A$ & Axin & nM & $A = \frac{\theta_6}{\mu_A}A'$ \\
\hline
$I_1$ &  Intermediate 1 (NICD/\bcat) & nM & $I_1 = \frac{\theta_1}{\mu_{I_1}}I_1'$\\
$I_2$ &  Intermediate 2 (\gsk{}/\bcat) & nM & $I_2 = \frac{\theta_1}{\mu_{I_2}}I_2'$ \\
\hline
\end{tabular}
\caption[]{Variable listings for our coupled Notch-Wnt ODE model. The independent variable, time, is presented first, followed by the twelve dependent variables representing network reactants.}
\label{tab: entities}
\end{table*}

\subsection*{Notch Pathway Submodel}
Our submodel for the Notch pathway is shown in Fig.~\ref{fig: NetworkDiagram}B, and includes both receptor-ligand binding at the cell surface membrane and subcellular details of Hes1 regulation. Its seven dependent variables are: Notch receptor, $N(t)$; Notch Intra-Cellular Domain (NICD), $F(t)$; intermediate $1$, $I_1(t)$, representing NICD bound to \bcat; Hes1, $H_1(t)$; proneural protein (Ngn3), $P(t)$; Delta ligand, $D(t)$; and Hath1 $H_2(t)$, as detailed in Table \ref{tab: entities}. 

The use of Hill-type and hyperbolic functional forms in the Notch submodel follows a four-component, non-delay model by Shepherd \cite{shepherd2010thesis}, itself an adaptation of a delay model due to Momiji and Monk \cite{momiji2009oscillatory}. Shepherd's model is capable of generating oscillations from a non-delay formulation.

\paragraph{Notch receptor, $N$.} A membrane-bound Delta ligand on a signalling cell can bind with a Notch receptor on a neighbouring cell \cite{vassin1987neurogenic}, initiating a series of reactions in the latter cell which constitute a \emph{Notch signalling cascade} \cite{artavanis1999notch}. Multiple forms of Delta have been identified, although it is not yet clear how their functions differ \cite{bray2006notch}. A further class of ligands, Jagged (Serrate in non-human mammals), is also complementary for Notch \cite{tax1994sequence}. For simplicity, we consider only Delta as the binding partner in our model; nor do we distinguish between the three distinct types of Delta ligand known to exist in mammals \cite{bray2006notch}.

We assume that the dominant processes regulating levels of Notch are its production and fragmentation. Production of Notch receptor is modelled by a Hill function, similar in form to the rates suggested by Shepherd \cite{shepherd2010thesis} and Collier \emph{et al.} \cite{collier1996pattern}. The linear decay term represents natural decay of Notch, as well as its fragmentation to release NICD:
\begin{eqnarray}
\label{eqn: notch}
 \frac{dN}{dt} = - \mu_N N + \underbrace{\frac{\theta_1 \bar{D}^{m_1}}{\kappa_1^{m_1} + \bar{D}^{m_1}}}_{\text{Steps \mycirc{1} and \mycirc{9}}}\,,
\end{eqnarray}
where $\bar{D}$ is the mean Delta level expressed by neighbouring cells and $\mu_N$ the rate constant of ligand fragmentation. The Hill function has
dissociation constant $\kappa_1$, maximal rate $\theta_1$ and Hill coefficient $m_1$.

\paragraph{Notch Intracellular Domain (NICD), $F$.} Delta-Notch binding induces three cleavage events in the Notch ligand, known as $S2$, $S3$ and $S4$ \cite{fortini2002gamma}. These cleavages at the cell membrane are in part mediated by an enzyme complex, $\gamma$-secretase, and cause the internal NICD fragment to be released into the cytoplasm \cite{es2005notch}. 

For simplicity, we treat the three cleavages as a single event. NICD is a fragment of the membrane-bound Notch receptor and so its production is assumed to scale with the rate of Notch fragmentation, as $\alpha_{\text{frag}}\mu_N N$. The nondimensional constant $\alpha_{\text{frag}}$ represents the proportion of Notch which results in generation of NICD; it is determined via parameter fitting against experimental data, as described in Methods and Models. $\alpha_{\text{frag}} < 1$ acknowledges that loss of Notch in Eqn. (\ref{eqn: notch}) is due to natural decay as well as NICD fragmentation. The rate of removal of NICD is assumed to depend on the rate at which it binds with \bcat, with rate constant $\alpha_1$, along with natural decay at rate $\mu_F$. These assumptions for production and loss of NICD yield the following ODE:
\begin{eqnarray}
\label{eqn: NICD}
 \frac{dF}{dt} = - \mu_F F + \underbrace{\alpha_{\text{frag}} \mu_N N}_{\text{Step \mycirc{2}}} - \underbrace{\alpha_1 B \cdot F}_{\text{Step \mycirc{3}}}\,.
\end{eqnarray}

\paragraph{Intermediate $1$, $I_1$.} Direct binding of NICD with \bcat{} is a key point of Notch-Wnt crosstalk. This interaction has been identified in vascular progenitor cells via reporter gene and immunoprecipitation studies, in which expression of NICD and/or \bcat{} were activated \cite{yamamizu2010convergence}. NICD-\bcat{} binding has also been demonstrated \invitro{} in human kidney cells \cite{jin2009beta}. NICD-mediated sequestering of \bcat{} may provide a switching mechanism in the canonical Wnt pathway by diverting \bcat{} from regulating Wnt target genes \cite{jin2009beta}.

Our model assumes the evolution of $I_1$ to be governed by its formation from \bcat{} and NICD, and its dissociation. In the absence of suitable experimental data, we assume a 1-1 stoichiometry for the binding of \bcat{} and NICD to form $I_1$ ($B + F \rightleftharpoons BF \, (= I_1)$). Given the strong experimental evidence for such binding, we assume that the rate at which the reverse reaction occurs is negligible. A first-order mass action law is then used to derive the reaction rate for intermediate formation, namely $\alpha_1 B\cdot F$. Dissociation of these intermediates adopts a simple linear form, with decay rate constant $\mu_{I_1}$. Combining these considerations yields the following ODE for $I_1$:
\begin{eqnarray}
\label{eqn: intermed-1}
 \frac{dI_1}{dt} = - \mu_{I_1} I_1 + \underbrace{\alpha_1 B \cdot F}_{\text{Step \mycirc{3}}}\,.
\end{eqnarray}

\paragraph{Hes1, $H_1$.} Following on from the evidence for Step \mycirc{3}, Yamamizu \etal{} \cite{yamamizu2010convergence} demonstrate a synergy between NICD and \bcat{} which amplifies the effect on the Hes1 promoter, Step \mycirc{4} in our model. A threefold increase in Hes1 promoter activity is observed in cells expressing both NICD and \bcat{}, compared to those expressing NICD alone \cite{yamamizu2010convergence}. This upregulation has also been observed in human kidney cells \cite{espinosa2003phosphorylation}, human colon cells \cite{peignon2011complex} and murine fibroblasts \cite{jin2009beta}. \emph{In vitro} studies of hamster ovarian cells identify Dishevelled as a key effector of Notch regulation at this point in the network, by measuring changes in Notch response when either Wnt, Dsh or \bcat{} production is induced \cite{collu2012dishevelled}. Expression of either Wnt or Dsh results in a $0.4-$ to $0.5-$fold change in Notch activity, yet expression of \bcat{} results in a $1.2-$fold change. Immunoprecipitation studies on the same cell line have confirmed the direct binding of Dsh to RBP-J$\kappa$ complexed with NICD. The same study also showed that blocking the Dsh signal allows human neuroblastoma lines to evade Notch downregulation in the face of a Wnt stimulus \cite{collu2012dishevelled}. Furthermore, Hes1 protein is known to dimerise and bind to its own promoter (Step \mycirc{5} in our model), thereby negatively regulating its own expression \cite{takebayashi1994structure, iso2003hes, kageyama2007hes}. This motif is highly conserved between species and is borne out by molecular analyses, \invitro{} and \invivo{} studies in mice \cite{takebayashi1994structure, ishibashi1995targeted, hirata2002oscillatory}, rats \cite{sasai1992two, hirata2002oscillatory} and \emph{Drosophila} \cite{takebayashi1994structure}, amongst others. This autoinhibition causes oscillations in the levels of Hes1 mRNA and protein and has been investigated by Hirata \etal{} \cite{hirata2002oscillatory}, who report an oscillation period of around two hours in a variety of cell lines and estimate the half-lives of the mRNA and protein to be around $24.1$ and $22.3$ minutes respectively.

In our model, Hes1 is subject to transcriptional regulation by $B$, $I_1$, $H_1$ and (implicitly) \emph{Dsh} and its transcription is assumed to have a maximal rate, $\xi_2$. $B$ and $I_1$ are assumed to be independent \emph{upregulators}, modelled via Hill functions with exponents $m_2, m_7$ and Hill coefficients $\kappa_2, \kappa_7$ (see Eqn. (\ref{eqn: hes1})), with an additive effect upon the promoter, owing to the separate promoter binding sites known to exist for $I_1$ and \bcat{} \cite{peignon2011complex}. The relative contributions of $I_1$ and $B$ to the upregulation are described by the non-negative, nondimensional constants $\theta_2, \theta_7$, such that $\theta_2 + \theta_7 = 1.0$. This constrains the sum of the two Hill functions to lie in the range $[0,1]$ and reflects our assumption that all upregulation of Hes1 is either Notch-mediated or Wnt-mediated.

\emph{Downregulators} of Hes1 transcription are $H_1$ (i.e. autorepression) and \emph{Dsh}. The autorepression is modelled by a hyperbola in $H_1$, with exponent $n_2$ and inhibition constant $\sigma_2$. For simplicity we do not represent Dsh explicitly in our model and assume instead that it is a decreasing function of Wnt, which we denote $\Psi_W$:
\begin{eqnarray*}
\label{eqn: sigmoid-W}
 \Psi_W = \Psi(W(t)) \equiv \frac{\sigma_K}{\sigma_K + W(t)}\,,
\end{eqnarray*}
where $\sigma_K$ is an inhibition constant to be specified. $\Psi_W$ attenuates the expression of Hes1 in response to a strong extracellular Wnt stimulus.

Combining these assumptions and assuming linear decay of Hes1 yields the following ODE for its evolution:
\begin{align}
\label{eqn: hes1}
 \frac{dH_1}{dt} = & -\mu_{H_1} H_1 + \Psi_W \Bigg( \underbrace{\frac{\theta_2 I_1^{m_2}}{\kappa_2^{m_2} + I_1^{m_2}}}_{\text{Step \mycirc{4}}} + \underbrace{\frac{\theta_7 B^{m_7}}{\kappa_7^{m_7} + B^{m_7}}}_{\text{Step \mycirc{14}}} \Bigg)\underbrace{\frac{\xi_2 \sigma_2^{n_2}}{\sigma_2^{n_2} + H_1^{n_2}}}_{\text{Step \mycirc{5}}}\,,
\end{align}
where $\mu_{H_1}$ is the rate of Hes1 decay.

\paragraph{Proneural protein, $P$.} Hes1 binds directly to the Ngn3 promoter, thereby blocking its activity \cite{jensen2000control}, although a destabilisation of Hes1 on Ngn3 has been postulated \cite{qu2013notch}. This downregulation has been observed \invitro{} in mouse fibroblasts \cite{lee2001regulation}, \invivo{} in the murine gut \cite{jensen2000control, lee2001regulation, fre2005notch}, and has been confirmed via sequence analyses of the Ngn3 promoter in humans and mice \cite{lee2001regulation}. Hes1-mutant mice embryos have been shown to exhibit premature neuronal differentation, associated with increased expression of the Neurogenin protein family \cite{ohtsuka1999hes1}; conversely, Ngn3-null mice exhibit diminished differentiation capacity in their intestinal cells \cite{jenny2002neurogenin3}. Hes1-knockout, adult mice show an elevated expression of Ngn3 in the gut \cite{ueo2012role}. Experiments on human liver cells estimate the half-life of Ngn3 to be around 30 
minutes \cite{qu2013notch}.

We account for the transcriptional inhibition of Ngn3 by Hes1, via a hyperbola in $H_1$, with maximal rate $\xi_3$, exponent $n_3$ and inhibition constant $\sigma_3$. If we assume further that Ngn3 undergoes natural decay, we obtain the following ODE for its evolution:
\begin{eqnarray}
\label{eqn: ngn3}
 \frac{dP}{dt} = -\mu_P P + \underbrace{\frac{\xi_3 \sigma_3^{n_3}}{\sigma_3^{n_3} + H_1^{n_3}}}_{\text{Step \mycirc{7}}} \,,
\end{eqnarray}
where $\mu_P$ is the rate of decay of proneural protein.

\paragraph{Delta ligand, $D$.} Upregulation of the Delta gene, $Dll1$, by proneural protein has been demonstrated in the mouse pancreas, through microarray analysis of Ngn3-inducible cell lines \cite{treff2006differentiation, gasa2004proendocrine} and coexpression studies \cite{qu2013notch}. This action appears to be highly conserved in non-mammalian species including \emph{Xenopus} \cite{seo2007neurogenin}. Steps \mycirc{7} and \mycirc{8} link the strength of the Notch signal inversely with that of Delta expression. Cells of the murine intestine which test null for Hes1 also have high Delta expression \cite{jensen2000control}; similarly, Math1(+/+) mice have low Hes1 levels and express more Delta \cite{yang2001requirement}.

Following Shepherd \cite{shepherd2010thesis}, we model the synthesis of Delta by a Hill function in $P$, with maximal rate $\theta_4$, Hill coefficient $m_4$ and dissociation constant $\kappa_4$. If we assume further that Delta undergoes natural decay at rate $\mu_D$ then we obtain the following ODE for its evolution:
\begin{eqnarray}
\label{eqn: delta}
 \frac{dD}{dt} = - \mu_D D + \underbrace{\frac{\theta_4 P^{m_4}}{\kappa_4^{m_4} + P^{m_4}}}_{\text{Step \mycirc{8}}}\,.
\end{eqnarray}

We note that, in a given cell, the reaction cascade running from Notch to Delta via Hes1 serves to downregulate Delta when that cell is expressing high levels of Notch. For this reason we expect either Step \mycirc{$1$} or Step \mycirc{$9$} to dominate in any one cell, in cases where the cascade is driving lateral inhibition or cell type segregation.

\paragraph{Hath1, $H_2$.} Gene expression analyses on the gut tissue of Hes1-knockout mice have revealed downregulation of Math1 by Hes1 \cite{jensen2000control, es2005bnotch, fre2005notch, ueo2012role}. These tissue samples show elevated numbers of secretory cells, typically associated with Hath1 expression \cite{jensen2000control, es2005bnotch}. Conversely, activation of Notch signalling (and hence Hes1 expression) in the murine gut \cite{fre2005notch} and in human CRC cell lines \cite{sikandar2010notch} represses the transcription of Math1/Hath1 and impairs the formation of goblet cell types. Whether this behaviour is more apparent in the early digestive tract \cite{jensen2000control} or prevalent throughout the gut \cite{es2005bnotch} remains to be established.

Accordingly, we use a hyperbola to model the inhibitory influence of Hes1 upon the production of Hath1, with maximal rate $\xi_5$, exponent $n_5$ and inhibition constant $\sigma_5$. If we assume further that Hath1 decays linearly with rate constant $\mu_{H_2}$, then we obtain the following ODE for $H_2$:
\begin{eqnarray}
\label{eqn: hath1}
 \frac{dH_2}{dt} = -\mu_{H_2} H_2 + \underbrace{\frac{\xi_5 \sigma_5^{n_5}}{\sigma_5^{n_5} + H_1^{n_5}}}_{\text{Step \mycirc{6}}}\,.
\end{eqnarray}
We note that $H_2$ decouples from the rest of the system; we retain Eqn. (\ref{eqn: hath1}) nonetheless, as it provides a read-out for cell fate specification.

\paragraph{} Eqns. (\ref{eqn: notch}) -- (\ref{eqn: hath1}) constitute our submodel for the Notch pathway. Crosstalk between the Notch and Wnt pathways centres upon NICD and Hes1, as we elucidate below when we describe our submodel for the Wnt pathway.

\subsection*{Wnt Pathway Submodel}
In the interest of focusing upon the dynamics surrounding the Hes1 crosstalk hub, we present a pared-down representation of the Wnt system. Our submodel depicted in Fig.~\ref{fig: NetworkDiagram}A comprises equations for \gsk, \bcat, Axin, a generalised ``destruction complex'' for \bcat{} and an intermediate formed from the aggregation of this complex with \bcat{} during the ubiquitination process of the \bcat{} destruction cycle. 

Although phosphorylated \bcat{} is formed during the dissociation of this intermediate, its evolution is assumed to have no bearing on the rest of the system, as the phosphorylated form is subsequently degraded by the proteasome, without participating in any other reactions. Consequently our model does not explicitly account for the evolution of phosphorylated \bcat.

\subsubsection*{Note}
The literature offers several detailed models of the Wnt pathway, e.g. \cite{lee2003roles, cho2006wnt, wawra2007extended}. Our focus on Notch-Wnt crosstalk motivates the use of a less detailed Wnt model. This aims to capture the qualitative behaviour of major Wnt pathway species, such as \gsk{} and \bcat, as demonstrated by the fitting of the \bcat{} concentration to data from Hern\'andez \etal{} in Fig.~\ref{fig: hernandez-matching}. 

In developing the Wnt submodel, analytic forms for the steady states of the Wnt species were determined using Eqns. (\ref{eqn: gsk}) -- (\ref{eqn: axin}). The functional form of terms for \bcat{} synthesis ($(1 + W(t))\alpha_{4}$), \gsk{} synthesis ($(1 + W(t))\alpha_2$) and formation of the destruction complex ($\Psi_{W,A}$) were determined by inspection of the steady state expressions. The Wnt-dependence of these terms was chosen such that the response of the steady states to changes in Wnt stimulation showed qualitative agreement with the literature (e.g. \bcat{} levels are enhanced by Wnt stimulation). Values for the exponents and multiplying parameter of $\Psi_{W,A}$ are determined by matching the Wnt response of the system to the experimental data of Hern\'andez \etal{} \cite{hernandez2012kinetic}, as described in \emph{Methods and Models - Parametrisation}. Future work might refine the Wnt submodel to incorporate other species and relax the Wnt-dependence of the synthesis terms.

\paragraph{\gsk, $G$.} When Wnt levels are low, \gsk{} is present in complexed form, and (along with the proteins Axin and APC) is a key component of a destruction complex that binds to \bcat{} and labels it for degradation \cite{bienz2000linking, fodde2007wnt}. When Wnt levels are high, the complex is largely disaggregated, yielding an increase in cytoplasmic \bcat.

In our model, we consider a general ``destruction complex'' rather than accounting for all of its component parts, and model the transition of \gsk{} between the complexed and non-complexed states as a reversible reaction, $G \rightleftharpoons C$. This may be an oversimplification, given assertions in the literature that the concentration of Axin within the complex may have a central role in regulating the rate of this step in \emph{Xenopus} oocytes \cite{lee2003roles}; however, \invitro{} experiments suggest that this is not the case for mammalian cell lines \cite{tan2012wnt}. Vesicular shuttling of \gsk{} is believed to play a role in coordinating Wnt signalling \cite{taelman2010wnt}, but our model does not distinguish between the various spatially sequestered forms of \gsk.

Owing to its dependence upon the local Wnt stimulus and the cellular Axin levels, we abbreviate the rate function for the forward reaction to $\Psi_{W,A}$. Applying the approach described in the above Note, we suppose the forward reaction $G \rightarrow C$ to be Wnt- and Axin-dependent, of rate
\begin{eqnarray}
\label{eqn: sigmoid-WA}
 \Psi_{W,A} = \Psi(W(t), A(t)) \equiv \frac{1.4 A(t)^{2}}{1 + (1 + W(t))^4}\,.
\end{eqnarray}
That is, the rate increases with Axin levels and decreases with Wnt stimulus. The reverse reaction $C \rightarrow G$ is assumed to occur at constant rate $\mu_C$. The constant $1.4$ is determined from the parameter fitting procedure described in the main text and in the above \emph{Note}.

In our model, the loss of \gsk{} arises from linear decay and transfer to the \bcat{} destruction complex, $C$. We assume that there is a basal rate, $\alpha_2$, of production of \gsk, and that this rate increases when there is a Wnt stimulus, $W(t)$. This is an artefact of having a small-scale Wnt model, being required to yield suitable steady-state behaviour in response to Wnt stimulation; refinement of this aspect of the model is a possible area for future work.

We assume further that \gsk{} binds reversibly with other proteins to form the destruction complex; these are assumed to be abundant. Combining these processes and assuming further that \gsk{} undergoes natural decay, we deduce that its evolution can be written as:
\begin{align}
\label{eqn: gsk}
 \frac{dG}{dt} = & - \mu_G G + (1 + W(t))\alpha_2 \underbrace{+ \mu_C C - \alpha_5 \Psi_{W,A}G}_{\text{Step \mycirc{13}}}\,,
\end{align}
where $\mu_G$ is the decay rate of $G$, $\mu_C$ the rate of dissociation of the destruction complex $C$ and $\alpha_5$ the rate constant for \gsk{} incorporation into the destruction complex.

\paragraph{Destruction Complex, $C$.}  
In our model, the destruction complex, $C$, is produced and lost at rates $\Psi_{W,A}$ and $\mu_C$ respectively, in the reverse manner to that of Eqn. (\ref{eqn: gsk}). We assume further that the destruction complex is released from the intermediate $I_2$ at rate $\mu_{I_2}$, and that it binds to \bcat{} at rate $\alpha_3$. Combining these processes, we deduce that the time evolution of $C$ can be written as:
\begin{eqnarray}
\label{eqn: dest-complex}
 \frac{dC}{dt} = \underbrace{-\mu_C C + \alpha_5 \Psi_{W,A}G}_{\text{Step \mycirc{13}}} + \mu_{I_2} I_2 - \underbrace{\alpha_3 B\cdot C}_{\text{Step \mycirc{11}}}\,, 
\end{eqnarray}
where $\Psi_{W,A}$ is defined as described above for \gsk{} in Eqn. (\ref{eqn: gsk}).

\paragraph{$\beta$-catenin, $B$.} Once \bcat{} is bound to the destruction complex, it is phosphorylated \cite{fodde2007wnt}, ubiquitinated and destroyed by the proteasome \cite{reya2005wnt}. Crosstalk between the Wnt and Notch pathways via \bcat{} is a key focus of our study. Peignon \etal{} \cite{peignon2011complex} have identified complementary binding sites on the \bcat{} molecule and Hes1 promoter and infer direct regulation of Hes1 levels by \bcat. This regulation occurs independently of the Notch-dependent mechanism described in Step \mycirc{4} above.

In our model, we assume that \bcat{} is produced at a basal rate, $\alpha_4$, which is enhanced when a Wnt stimulus is present. Loss of \bcat{} arises from its binding to either NICD or the destruction complex, at rates $\alpha_1$ and $\alpha_3$ respectively, to form the intermediates $I_1, I_2$; natural decay of \bcat{} is also assumed, at rate $\mu_B$. Combining these assumptions, we obtain the following evolution equation for \bcat:
\begin{align}
\label{eqn: bcat}
 \frac{dB}{dt} = & -\mu_B B + (1 + W(t))\alpha_{4} - \underbrace{\alpha_1 B \cdot F}_{\text{Step \mycirc{3}}} - \underbrace{\alpha_3 B \cdot C}_{\text{Step \mycirc{14}}}\,.
\end{align}

\paragraph{Intermediate $2$, $I_2$.} Intermediate $2$ arises from the binding of the destruction complex to \bcat. As for intermediate $I_1$, we assume that the forward reaction dominates and follows first-order mass action, whilst the dissociation of $I_2$ has rate constant $\mu_{I_2}$. These assumptions yield the following ODE for $I_2$:
\begin{eqnarray}
\label{eqn: intermed-2}
 \frac{dI_2}{dt} = -\mu_{I_2} I_2 + \underbrace{\alpha_3 B \cdot C}_{\text{Step \mycirc{11}}}\,.
\end{eqnarray}

\paragraph{Axin, $A$.} The scaffold protein Axin plays a key role in the action of the destruction complex on \bcat{} \cite{jho2002wnt, lustig2003wnt}. It is also transcriptionally upregulated by \bcat{} and, as such, forms a negative feedback loop, serving to regulate the Wnt pathway \cite{jho2002wnt}. Induction of Axin in response to Wnt exposure has been documented in cell lines from mice \cite{jho2002wnt}, rats \cite{leung2002activation, jho2002wnt} and human colon cancer \cite{leung2002activation}.

We model Axin synthesis using a Hill function which depends upon $B$, with maximal rate $\theta_6$, Hill coefficient $m_6$ and dissociation constant $\kappa_6$. Assuming further that Axin decays linearly at rate $\mu_A$, we have:
\begin{eqnarray}
 \label{eqn: axin}
 \frac{dA}{dt} = -\mu_A A + \underbrace{\frac{\theta_6 B^{m_6}}{\kappa_6^{m_6} + B^{m_6}}}_{\text{Step \mycirc{10}}}\,.
\end{eqnarray}
Axin presents in two functionally equivalent forms, Axin1 and Axin2, but we do not differentiate between these in our model. Both form part of the \bcat{} destruction complex, but result in different phenotypes when deleted in mice \cite{mazzoni2014axin1}. There is sufficient scope to develop the model to account for these forms: for example, the main Axin variable could be taken to represent Axin2, and the influence of Axin1 could be incorporated via dependencies for the dissociation constants of the Hill functions in either \bcat{} (Eqn. (\ref{eqn: bcat})) or Axin (Eqn. (\ref{eqn: axin})). We note this as an area for development of the model in future.

Initial conditions for the dimensional model are specified in Table \ref{tab: initial-conditions}.

\subsection*{Nondimensionalised System}
\label{sec: SI-nondim}
Eqns. (\ref{eqn: notch}) - (\ref{eqn: axin}) define our ODE system in dimensional form; we shall now nondimensionalise it to reduce the number of system parameters and to facilitate estimation of the relative importance of the different reactions within the network. 

Time is scaled against $\mu_N^{-1}$, the timescale for decay of the Notch receptor; we define a nondimensional time, $\tau$, such that $\tau = \mu_N t$. Scalings for dependent variables are introduced in Table \ref{tab: entities}, and those for parameters in Table \ref{tab: parameters}. For reactant $X$, we also introduce the dimensionless parameter $\nu_X = \mu_X/\mu_N$ to represent the ratio of its decay rate to that of Notch. The twelve ODEs for the nondimensional system are as follows (primes denote dimensionless variables):

\begin{align}
\label{eqn: notch-nondim}
 \frac{dN'}{d\tau} &= -N' + \frac{\bar{D'}^{m_1}}{\kappa_1'^{m_1} + \bar{D'}^{m_1}},\qquad \\
\label{eqn: NICD-nondim}
 \frac{dF'}{d\tau} &= \nu_F \Bigg\{ -F' + \alpha_{\text{frag}} N' - \frac{\alpha_1'}{\nu_F}B'\cdot F' \Bigg\} \,, \qquad \qquad \qquad \\
\label{eqn: intermed-1-nondim}
 \frac{dI_1'}{d\tau} &= \nu_{I_1} \Bigg\{ -I_1' + \frac{\alpha_1'}{\nu_F}B'\cdot F' \Bigg\} \,,\qquad \qquad \\
\label{eqn: hes1-nondim}
 \frac{dH_1'}{d\tau} &= \nu_{H_1} \Bigg\{ -H_1' + \Psi_W \Bigg(\frac{\theta_2 I_1'^{m_2}}{\kappa_2'^{m_2} + I_1'^{m_2}} + \frac{\theta_7 B'^{m_7}}{\kappa_7'^{m_7} + B'^{m_7}} \Bigg)\frac{\sigma_2'^{n_2}}{\sigma_2'^{n_2} + H_1'^{n_2}} \Bigg\}\,, \\
 \label{eqn: ngn3-nondim}
 \frac{dP'}{d\tau} &= \nu_P \Bigg\{ -P' + \frac{\sigma_3'^{n_3}}{\sigma_3'^{n_3} + H_1'^{n_3}} \Bigg\} \,,\qquad \qquad \\
 \label{eqn: delta-nondim}
 \frac{dD'}{d\tau} &= \nu_D \Bigg\{ -D' + \frac{P'^{m_4}}{\kappa_4'^{m_4} + P'^{m_4}} \Bigg\} \,,\qquad \qquad \\
\label{eqn: hath1-nondim}
 \frac{dH_2'}{d\tau} &= \nu_{H_2} \Bigg\{ -H_2' + \frac{\sigma_5'^{n_5}}{\sigma_5'^{n_5} + H_1'^{n_5}} \Bigg\} \,, \qquad \\
 \label{eqn: gsk-nondim}
 \frac{dG'}{d\tau} &=  \nu_G \Bigg\{ -G' + C' + (1 + W(\tau))\alpha_2' - \alpha_5'\Psi_{W,A'}G' \Bigg\} \,,\\
\label{eqn: dest-complex-nondim}
 \frac{dC'}{d\tau} &= \nu_C \Bigg\{ -C' + I_2' + \alpha_5'\Psi_{W,A'}G' -\frac{\alpha_3'}{\nu_C}B'\cdot C' \Bigg\},\\
\label{eqn: bcat-nondim}
 \frac{dB'}{d\tau} &= \nu_B \Bigg\{- B' + (1 + W(\tau))\alpha_4' - \frac{\alpha_1'}{\nu_F}B'\cdot F' - \frac{\alpha_3'}{\nu_C}B'\cdot C' \Bigg\} \,,\\
 \label{eqn: intermed-2-nondim}
 \frac{dI_2'}{d\tau} &= \nu_{I_2} \Bigg\{ -I_2' + \frac{\alpha_3'}{\nu_C}B'\cdot C' \Bigg\} \,, \qquad \quad \qquad \\
\label{eqn: axin-nondim}
 \frac{dA'}{d\tau} &= \nu_A \Bigg\{ -A' + \frac{B'^{m_6}}{\kappa_6'^{m_6} + B'^{m_6}} \Bigg\}\,.
\end{align}
For brevity, we hereafter drop the prime notation from our nondimensional variables.

\begin{table*}[t]
\centering
\renewcommand{\arraystretch}{1.5}
\begin{tabular}{|c|c|c|c|}
\hline
& \textbf{Parameter} & \textbf{Units} & \textbf{Scaling}\\
\hline
\multirow{5}{*}{\textbf{\emph{Dissociation Constants}}}&$\kappa_1$	& nM & $\kappa_1 = \frac{\theta_4}{\mu_D}\kappa_1'$ \\
& $\kappa_2$	& nM & $\kappa_2 = \frac{\theta_1}{\mu_{I_1}}\kappa_2'$\\
& $\kappa_4$	& nM & $\kappa_4 = \frac{\xi_3}{\mu_P}\kappa_4'$\\
& $\kappa_6$	& nM & $\kappa_6 = \frac{\theta_1}{\mu_N}\kappa_6'$\\
& $\kappa_7$	& nM & $\kappa_7 = \frac{\theta_1}{\mu_N}\kappa_7'$\\
\hline
\multirow{5}{*}{\textbf{\emph{Inhibition Constants}}}& $\sigma_2$	& nM & $\sigma_2 = \frac{\xi_2}{\mu_{H_1}}\sigma_2'$\\
& $\sigma_3$	& nM & $\sigma_3 = \frac{\xi_2}{\mu_{H_1}}\sigma_3'$\\
& $\sigma_5$	& nM & $\sigma_5 = \frac{\xi_2}{\mu_{H_1}}\sigma_5'$\\
& $\sigma_K$	& dim'less & - \\
\hline
\multirow{1}{*}{\textbf{\emph{Decay Rates}}} & General $\mu_X$ & min$^{-1}$ & $\mu_X = \nu_X \mu_N$\\
\hline
\multirow{1}{*}{\textbf{\emph{Decay Ratios}}} & General $\nu_X = \frac{\mu_X}{\mu_N}$ & dim'less & - \\
\hline
\multirow{3}{*}{\textbf{\emph{Maximal Values}}} & $\theta_i,\,(i=1,4,6)$ & nM min$^{-1}$ & - \\
& $\theta_i,\,(i=2,7)$ & dim'less & - \\
& $\xi_i,\,(i=2,3,5)$ & nM min$^{-1}$ & - \\
\hline
\multirow{5}{*}{\textbf{\emph{Other Constants}}} &$\alpha_1$ & nM$^{-1}$min$^{-1}$  & $\alpha_1 = \frac{\mu_N^2}{\theta_1} \alpha_1'$ \\
&$\alpha_2$ & nM min$^{-1}$  & $\alpha_2 = \theta_1\alpha_2'$ \\
&$\alpha_3$ & nM$^{-1}$ min$^{-1}$ & $\alpha_3 = \frac{\mu_N^2}{\theta_1}\alpha_3'$ \\
&$\alpha_4$ & nM min$^{-1}$ & $\alpha_4 = \theta_1\alpha_4'$ \\
&$\alpha_5$ & nM$^{-1}$ min$^{-1}$ & $\alpha_5 = \frac{\mu_G \mu_A}{\theta_6}\alpha_5'$\\
& $\alpha_{\text{frag}}$ & dim'less & - \\
\hline
\multirow{2}{*}{\textbf{\emph{Exponents}}} & $m_i,\,i=1,2,4,6,7$ & dim'less & - \\
&$n_i,\,i=2,3,5$ & dim'less & - \\
\hline
\end{tabular}
\caption[]{Parameters associated with the coupled Notch-Wnt ODE model presented in Eqns. (\ref{eqn: notch}) -- (\ref{eqn: axin})}
\label{tab: parameters}
\end{table*}

\subsection*{Model Parameters}
\label{sec: parametrisation}
Dimensional values for all parameters, along with supporting references, are shown in the following tables (listed in full at end of document):
\begin{itemize}
 \item \textbf{Half-lives and decay rates}: Table \ref{tab: dimensional-decay-rates};
 \item \textbf{Hill parameters}: Table \ref{tab: dimensional-hill-params};
 \item \textbf{Hyperbola parameters}: Table \ref{tab: dimensional-hyperbola-params};
 \item \textbf{Other parameters}: Table \ref{tab: general-rate-params};
 \item \textbf{Initial conditions}: Table \ref{tab: initial-conditions}.
\end{itemize}
Suitable experimental data are not currently available to estimate values of all model parameters. Where appropriate data were lacking, parameter estimation was performed as described in Methods and Models (Parametrisation); numerically fitted values are indicated by a `\emph{PF}' and/or footnotes.

\begin{sidewaystable}[p]
\centering
\renewcommand{\arraystretch}{1.5}
\begin{tabular}{|c|c|c|c|}
\hline
\textbf{Parameter} & \textbf{Half-Life (min)} & \textbf{Source} & \textbf{Estimated Value (min$^{-1}$)} \\
\hline
$\mu_N$ & $40.0$ & Logeat \etal{} \cite{logeat1998notch1}, human CRC cells & $0.017$ \\
$\mu_F$ & $180.0$ & Foltz \etal{} \cite{foltz2002glycogen}, murine fibroblasts & $0.00385$ \\
$\mu_{I_1}$ & - & \emph{PF} & $0.04$ \\
$\mu_{H_1}$ & $22.3$ & Hirata \etal{} \cite{hirata2002oscillatory}, murine myoblasts & $0.065$\footnote{Value derived by parameter fitting, using the estimate of $0.0311$ as a starting value.} \\
$\mu_P$ & $14.4$ & Roark \etal{} \cite{roark2012complex}, \emph{Xenopus laevis} & $0.035$\footnote{Value derived by parameter fitting, using the estimate of $0.0481$ as a starting value.} \\
$\mu_D$ & - & Collier \etal{} \cite{collier1996pattern}, Logeat \etal{} \cite{logeat1998notch1} & $0.049$\footnote{Value derived by parameter fitting from the Notch data of Logeat \etal{} \cite{logeat1998notch1}, using $\mu_D = \mu_N$ as in the model of Collier \etal{} \cite{collier1996pattern}.} \\
$\mu_{H_2}$ & - & \emph{PF} & $0.0311$\footnote{In the absence of data for $\mu_{H_2}$, a starting value was based upon that for $\mu_{H_1}$ from the data of Hirata \etal{} \cite{hirata2002oscillatory}.} \\
\hline
$\mu_G$ & 900.0 & Cole \etal{} \cite{cole2004further}, $HEK293$ cells & $9.36 \times 10^{-4,}$\footnote{Value derived by parameter fitting, using the estimate of $7.7 \times 10^{-4}$ as a starting value.} \\
$\mu_C$ & - & Lee \etal{} \cite{lee2003roles}, \emph{Xenopus} oocytes & $1.061$\footnote{Value derived by parameter fitting, using the estimate of $0.909$ as a starting value.} \\
$\mu_B$ & 104.0 & Hernandez \etal{} \cite{hernandez2012kinetic}, $RKO$ cells  & $0.00636$\footnote{Value derived by parameter fitting, using the estimate of $0.00666$ as a starting value.} \\
$\mu_A$ & 480.0 & Yamamoto \etal{} \cite{yamamoto1999phosphorylation}, $COS$ cells & $6.23 \times 10^{-4,}$\footnote{Value derived by parameter fitting, using the estimate of $0.00144$ as a starting value.} \\
$\mu_{I_2}$ & - & Lee \etal{} \cite{lee2003roles}, \emph{Xenopus} oocytes & $4.204$\footnote{Value derived by parameter fitting, using the estimate of $210.0$ as a starting value.}  \\
\hline
\end{tabular}
\caption[]{Dimensional estimates for the decay rate constants of Eqns. (\ref{eqn: notch}) -- (\ref{eqn: axin}). In the absence of suitable data, our estimates of $\mu_{H_2}$ and $\mu_{D}$ have been estimated using other known decay rates. Such approximations are based on factors such as the molecular size of a species and its location and function within the cell. \emph{PF} indicates numerically fitted parameters.}
\label{tab: dimensional-decay-rates}
\end{sidewaystable}

\begin{sidewaystable}[p]
\centering
\renewcommand{\arraystretch}{1.5}
\begin{tabular}{|c|c|c|c|}
\hline
\textbf{Parameter} & \textbf{Units} & \textbf{Dimensional Value} & \textbf{Source} \\
\hline
$\theta_1$ & nM min$^{-1}$ & $0.06$\footnote{Value derived by parameter fitting, using the estimate of $0.055$ from the calculations of Agrawal \etal{} \cite{agrawal2009computational} as an initial estimate.} & Agrawal \etal{} \cite{agrawal2009computational}\\
$\theta_2$	& dim'less & $0.75$  & Agrawal \etal{} \cite{agrawal2009computational}, Cinquin \etal{} \cite{cinquin2007repressor}\\
$\theta_4$	& nM min$^{-1}$ & $0.4$ & \emph{PF} \\
$\theta_6$	& nM min$^{-1}$ & $1.64 \times 10^{-4,}$\footnote{Value derived by parameter fitting, using the estimate of $8.2 \times 10^{-5}$ as a starting value.} & Lee \etal{} \cite{lee2003roles} \\
$\theta_7$	& dim'less & $0.25$  & Agrawal \etal{} \cite{agrawal2009computational}, Cinquin \etal{} \cite{cinquin2007repressor}\\
\hline
$\kappa_1$	& nM & $0.6$  & \emph{PF} \\
$\kappa_2$	& nM & $0.9$  & \emph{PF} \\
$\kappa_4$	& nM & $14.7$  & \emph{PF} \\
$\kappa_6$	& nM & $0.026$  & \emph{PF} \\
$\kappa_7$	& nM & $10.0$  & \emph{PF} \\
\hline
$m_i, i=1,2,4,6,7$	& dim'less & $3$ & - \\
\hline
\end{tabular}
\caption[]{Dimensional parameters associated with the Hill functions used in Eqns. (\ref{eqn: notch}) -- (\ref{eqn: axin}). The $\theta_i$ are maximal terms; the $\kappa_i$ are the Hill coefficients; and the $m_i$ are the Hill exponents. \emph{PF} indicates numerically fitted parameters.}
\label{tab: dimensional-hill-params}
\end{sidewaystable}

\begin{table}[t]
\centering
\renewcommand{\arraystretch}{1.5}
\begin{tabular}{|c|c|c|c|}
\hline
\textbf{Parameter} & \textbf{Units} & \textbf{Dimensional Value} & \textbf{Source}\\
\hline
$\xi_2$	& nM min$^{-1}$ & $0.5$  & Agrawal \etal{} \cite{agrawal2009computational} \\
$\xi_3$	& nM min$^{-1}$ & $0.9$  & \emph{PF} \\
$\xi_5$	& nM min$^{-1}$ & $0.9$  & \emph{PF} \\
\hline
$\sigma_2$	& nM & $3.5$  & \emph{PF} \\
$\sigma_3$	& nM & $1.21$  & \emph{PF} \\
$\sigma_5$	& nM & $1.7$  & \emph{PF} \\
\hline
$n_i, (i=2,3,5)$ & dim'less & $3$ & - \\
\hline
\end{tabular}
\caption[]{Dimensional parameters for the hyperbola functions used in Eqns. (\ref{eqn: notch}) -- (\ref{eqn: axin}). The $\xi_i$ are maximal rates; the $\sigma_i$ are constants of inhibition; and the $n_i$ are the exponents. \emph{PF} indicates numerically fitted parameters.}
\label{tab: dimensional-hyperbola-params}
\end{table}

\begin{table}[t]
\centering
\renewcommand{\arraystretch}{1.5}
\begin{tabular}{|c|c|c|c|}
\hline
\textbf{Parameter} & \textbf{Units} & \textbf{Dimensional Value} & \textbf{Source}\\
\hline
$\alpha_1$	& nM$^{-1}$ min$^{-1}$ & $6.8$ & \emph{PF} \\
$\alpha_2$	& nM min$^{-1}$ & $0.0174$ & \emph{PF}\footnote{In the absence of any experimental data, a starting value of $0.01$ was used in parameter fitting} \\
$\alpha_3$	& nM$^{-1}$ min$^{-1}$ & $1.465 \times 10^{-4}$& \emph{PF}\footnote{Value derived by parameter fitting against the \bcat{} timecourses of Hernandez \etal{} \cite{hernandez2012kinetic}, initial estimate $1.028 \times 10^{-4}$} \\
$\alpha_4$	& nM min$^{-1}$ & $0.472$ & \emph{PF}\footnote{Value derived by parameter fitting against the \bcat{} timecourses of Hernandez \etal{} \cite{hernandez2012kinetic}, initial estimate $0.3$} \\
$\alpha_5$	& nM$^{-1}$ min$^{-1}$ & $0.1044$ & \emph{PF}\footnote{Starting value of $0.091$, from the model of Lee \etal{} \cite{lee2003roles}}\\
\hline
$\sigma_K$	& dim'less & 1.0 & Collu \etal{} \cite{collu2012dishevelled} \\
$\alpha_{\text{frag}}$ & dim'less & $0.8$ & \emph{PF} \\
\hline
\end{tabular}
\caption[]{Miscellaneous rate parameters used in Eqns. (\ref{eqn: notch}) -- (\ref{eqn: axin}). \emph{PF} indicates numerically fitted parameters.}
\label{tab: general-rate-params}
\end{table}

\begin{table}[t]
\centering
\renewcommand{\arraystretch}{1.5}
\begin{tabular}{|c|c|c|}
\hline
\textbf{Variable} & \textbf{Description} & \textbf{Two-cell ICs} \\
& & \textbf{(nM)} \\
\hline
$N$	& Membrane-bound Notch receptor & $0.5$ \\
$F$ &  Notch Intracellular Domain (NICD) & $0.5$ \\
$H_1$ &  Hes1 & $0.5$ \\
$P$ & Ngn3 & $0.5$ \\
$D$	& Delta ligand & $0.5$ \\
$H_2$ &  Hath1 & $0.5$ \\
\hline
$G$ & \gsk & $30.0$ \\
$C$ & Destruction complex & $25.0$ \\
$B$ & Active \bcat & $9.0$ \\
$A$ & Axin & $27.0$ \\
\hline
$I_1$ &  Intermediate 1 (NICD/\bcat) & $0.5$ \\
$I_2$ &  Intermediate 2 (\gsk{}/\bcat) & $30.0$  \\
\hline
\end{tabular}
\caption[]{Initial conditions (ICs) for the variables in the dimensional Notch-Wnt ODE model.}
\label{tab: initial-conditions}
\end{table}

\subsection*{Steady-state analysis}

\subsubsection*{Analysis of Notch submodel}
Decoupling Notch from the Wnt system leaves the system shown in Fig.~\ref{fig: NetworkDiagram}B, represented by the ODEs (\ref{eqn: notch-nondim}) -- (\ref{eqn: hath1-nondim}). For clarity, we abbreviate the Hill functions and hyperbolas as follows:
\begin{eqnarray*}
 \Phi_i(X) = \frac{X^{m_i}}{\kappa_i^{m_i} + X^{m_i}}, \quad \Theta_i(X) = \frac{\sigma_i^{n_i}}{\sigma_i^{n_i} + X^{n_i}} \,,
\end{eqnarray*}
where all parameters are defined as in Table \ref{tab: parameters} and $X \equiv X(\tau)$ is a reactant concentration at time $\tau$. In what follows, we exploit the fact that $\Phi$ is monotonic increasing and $\Theta$ monotonic decreasing on $\mathbb{R}^+$, and hence that $\frac{d\Theta(X)}{dX}<0<\frac{d\Phi(X)}{dX}$, $\forall X \in \mathbb{R}^+$. To simplify the analysis further, we initially restrict attention to the homogeneous case, for which the cell population evolves in a uniform state. Consequently average $\bar{D}$ in Eqn. (\ref{eqn: delta}) is replaced by $D$. 

$B$ is treated as an input parameter throughout the Notch-only analysis, enabling us to exploit linearity within equations (\ref{eqn: NICD-nondim}) and (\ref{eqn: intermed-1-nondim}) to create a four-equation system for $N$, $H_1$, $P$ and $D$. Application of a steady state assumption ultimately yields the following implicit expression for the steady state concentration of Hes1, $H_1 = H^*$:
\begin{align}
\label{eqn: hes1-stst}
 H^* &= \Psi_W \Big(\theta_2 \Phi_2(\eta \Phi_1 \circ \Phi_4 \circ \Theta_3(H^*)) + \theta_7 \Phi_7(B)\Big)\Theta_2(H^*) \,.
\end{align}
Since $\Phi_i$ is an increasing and $\Theta_i$ a decreasing function for $H^* \in \mathbb{R}^+$,$\forall i$, the composition $\Phi_2(\eta
\Phi_1 \circ \Phi_4 \circ \Theta_3(H^*))$ is decreasing in $\mathbb{R}^+$. The right-hand side of (\ref{eqn: hes1-stst}) is therefore a decreasing,
positive-valued function of $H^*$ on $\mathbb{R}^+$, because $\Phi(x),\Theta(x) > 0$ for $x > 0$. Equality with the left-hand side, which is trivially
increasing in $H^*$, guarantees the existence of a unique biologically realistic steady state for Hes1.

\begin{figure}[htp]
\centering
\includegraphics[width=\textwidth]{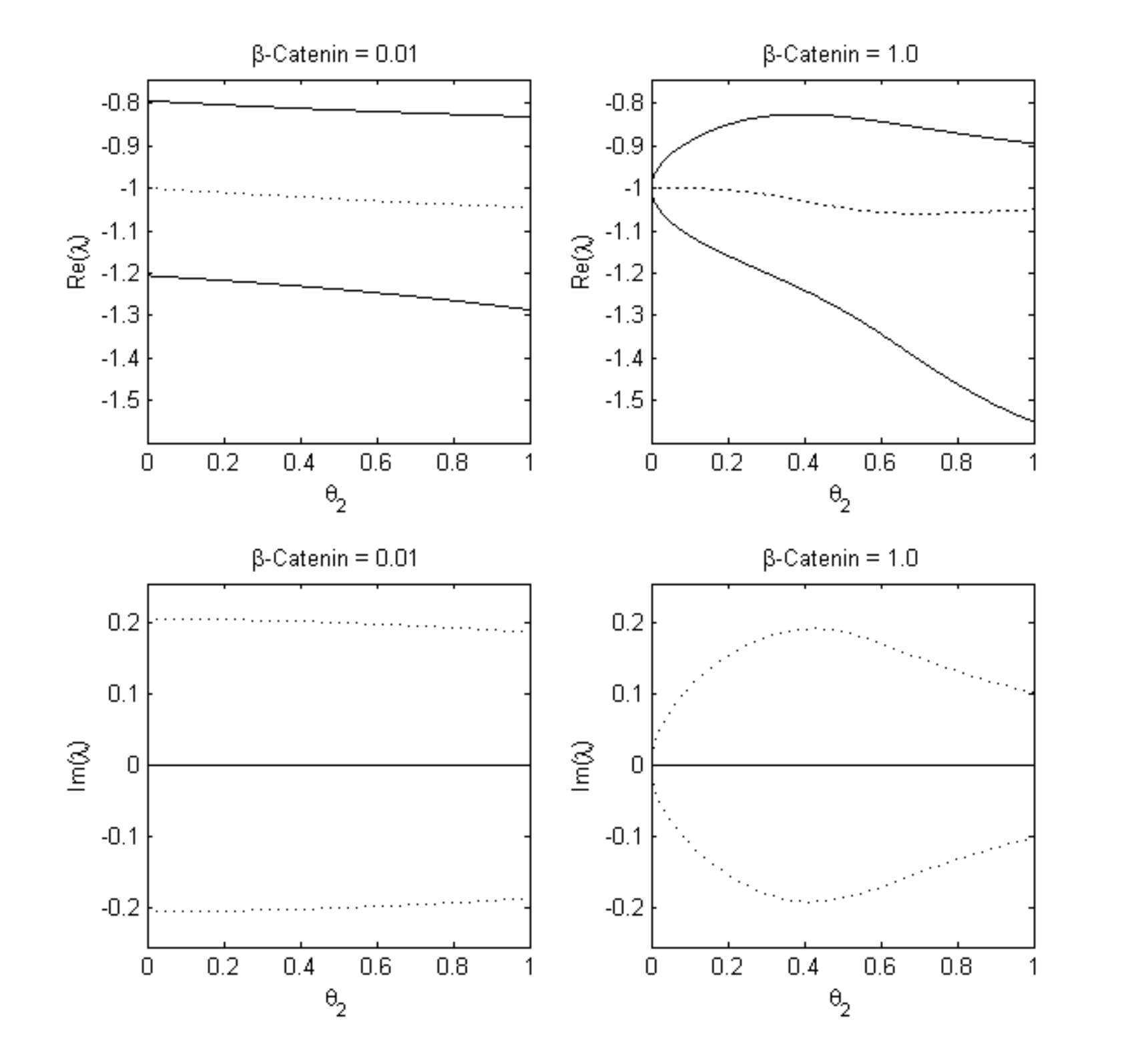}
\caption[]{{\bf{Eigenvalue dependence in the decoupled Notch system}} Dependence of the eigenvalues of the four-component ($N$, $H_1$, $P$, $D$) Notch system upon $\theta_2$, calculated using Eqn. (\ref{eqn: hes1-stst}). Plots depict (upper plots) real and (lower plots) imaginary components. Complex conjugate eigenvalues are depicted with dashed lines and we have $\nu_H = \nu_P = \nu_D = 1.0$; $\Psi_W = 1.0$; $\eta = 0.5$. Parameter values: $m_i, n_i = 3$; $\kappa_i = 0.5$ for $i = 1,2,4$; $\sigma_2 = 0.5$; $\sigma_3 = 0.1$.}
\label{fig: NotchEigenvalues}
\end{figure}

We now perform linear stability analysis of the steady-state solutions. Eigenvalue analysis of the four-component system suggests the existence of two negative reals, and a pair of complex conjugates with negative real part, as shown in Fig.~\ref{fig: NotchEigenvalues}. Negativity of the real parts of all four eigenvalues generates stable steady states in the system. The complex conjugate eigenvalues arise from a supercritical Hopf bifurcation in $\theta_2$, visible on the lower-right plot of Fig.~\ref{fig: NotchEigenvalues}. That this occurs for $\theta_2=0$ suggests that the action of \bcat{} upon the Hes1 promoter serves to stabilise Notch and dampen oscillations. Model simulations over a wide range of initial conditions (data not shown) indicate that at lower levels of $B$, Hes1 steady states increase with the amount of Notch-mediated transcriptional control (i.e. high $\theta_2$ values). Sufficiently high \bcat{} expression forces Hes1 into a lower steady state and can dominate Notch-mediated promotion of Hes1.

\subsubsection*{Analysis of Wnt submodel}
Our decoupled Wnt system can be described by Eqns. (\ref{eqn: gsk-nondim}) -- (\ref{eqn: axin-nondim}), representing the network depicted in Fig.~\ref{fig: NetworkDiagram}A. The concentration of NICD ($F$), which interacts with \bcat, is treated as a model parameter and a constant Wnt stimulus, $W$, is assumed. An implicit equation for the steady state $B^* = B^*(W,F)$ of \bcat{} is as follows:
\begin{align}
\label{eqn: bcat-stst}
 \frac{\alpha_4}{B^*} = \frac{1}{1 + W}\Bigg(1 + \frac{\alpha_1}{\nu_F}F \Bigg) + \frac{1.4\,\alpha_2 \alpha_3 \alpha_5}{\nu_C} \frac{1}{1 + (1 + W)^4} \, \Phi_6(B^*)^2.
\end{align}
Since the left-hand side of Eqn. (\ref{eqn: bcat-stst}) is decreasing in $B^*$ and the right-hand side is increasing in $B^*$, we deduce that there is a unique biologically realistic (i.e. in $\mathbb{R}^+$) solution for $B^*$. Inspection of Eqn. (\ref{eqn: bcat-stst}) confirms that $B^*$ exhibits qualitatively appropriate behaviour within the Wnt system. For example, the steady state value $B^*$ will be increased by:
\begin{itemize} \setlength{\itemsep}{-3pt}
 \item increasing the Wnt stimulus, $W(\tau)$;
 \item increasing $\alpha_4$, the production rate of \bcat{};
 \item decreasing $\alpha_5$, the rate constant for formation of the destruction complex, $C$;
 \item decreasing $\alpha_3$, the rate at which $C$ binds with \bcat{} ;
 \item decreasing the interaction of \bcat{} with the Notch system (involving elements of the group $\frac{\alpha_1}{\nu_C}F$).
\end{itemize}

\begin{figure}[tb]
\centering
\includegraphics[width=0.8\textwidth]{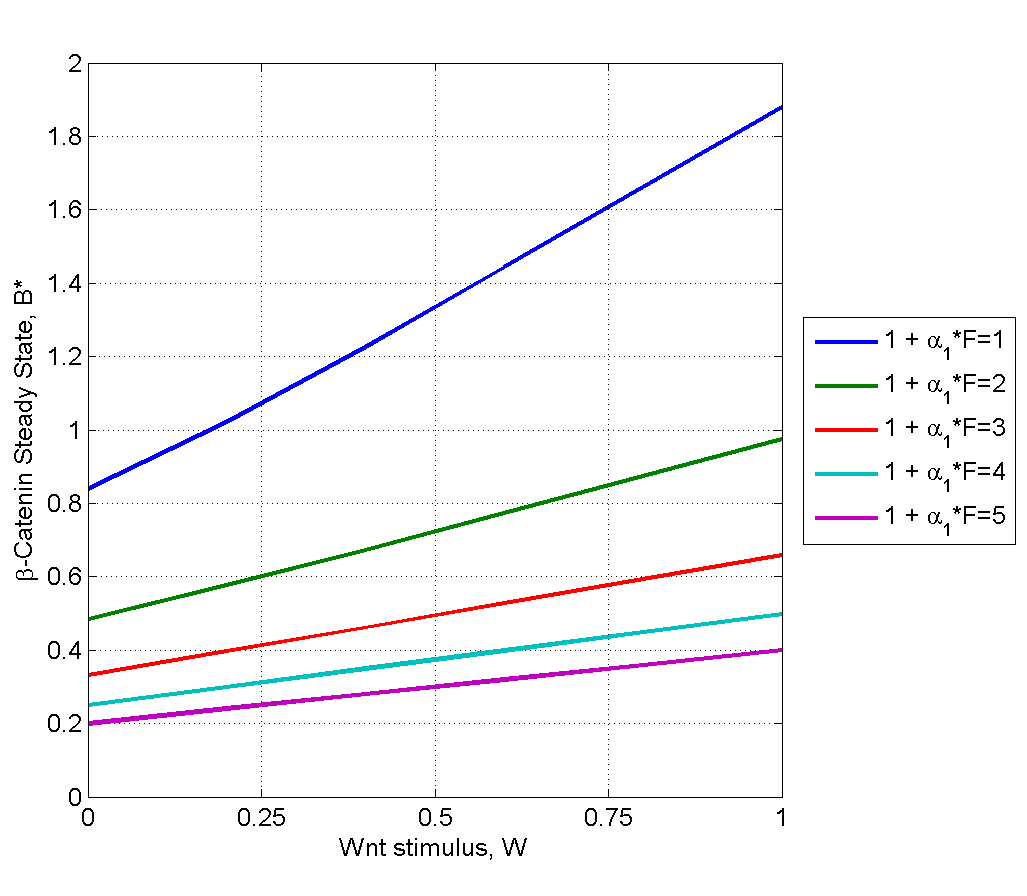}
\caption[]{{\bf{\bcat{} steady state response in the decoupled Wnt system}} Response of the \bcat{} steady state, $B^*$, described implicitly in Eqn. (\ref{eqn: bcat-stst}), to a variation in Wnt signal $W$ and strength of the Notch interaction, $\alpha_1 F$. Numerical solutions of (\ref{eqn: bcat-stst}) were generated from parameter values $\alpha_2 = \alpha_3 = \alpha_4 = 1.0$, $\alpha_5 = 0.4$ and $\nu_F = \nu_C = 1.0$. In our model, the stronger the interaction with the Notch system (determined by $\alpha_1 F$), the shallower the gradient of $B^*$ and hence the weaker the response of the Wnt system to variation in the extracellular Wnt stimulus.}
\label{fig: WntOnlyCrosstalk}
\end{figure}

Strong interaction with NICD attenuates the response of \bcat{} to variation in Wnt levels, as shown in Fig.~\ref{fig: WntOnlyCrosstalk}. As the quantity $\alpha_1 F$ increases, the gradient of $B^*$ with respect to Wnt stimulus tends to zero. When the crosstalk with Notch is reduced, $B^*$ increases with the Wnt signal.

\subsection*{A Formal Framework of CRNT}
The following definitions and examples aim to provide a brief introduction to the concordance property. A full treatment can be found in \cite{shinar2012concordant,shinar2013concordant}. Except where stated otherwise, our definitions and propositions are adopted from Shinar and Feinberg \cite{shinar2012concordant}.

A biochemical network consists of two aspects: a \emph{network structure} $\{\mathcal{S,C,R}\}$, where:
\begin{itemize}
 \item $\mathcal{S}$ is the set of all \emph{chemical species} in the network, $\mathcal{S}=\{X_1, X_2, \ldots, X_n\}$;
 \item $\mathcal{C}$ is the set of \emph{complexes}. In CRNT terminology, $\mathcal{C}$ comprises reactant or product expressions which express the stoichiometric linking of elements of $\mathcal{S}$, such as $F + B$, $N$, and $B + C$ -- i.e., any expression which forms the entire right- or left-hand side of a reaction equation;
 \item $\mathcal{R} \subset \mathcal{C} \times \mathcal{C}$ is the set of \emph{reactions}. Reactions satisfy two properties: first, that $(y,y) \not\in \mathcal{R}$ for any $y \in \mathcal{C}$; secondly, that for each $y \in \mathcal{C}$, there exists $y' \in \mathcal{C}$ such that $(y,y') \in \mathcal{R}$ or $(y',y) \in \mathcal{R}$;
\end{itemize}
and a \emph{kinetics}, $\mathcal{K}$, which assigns a functional form to each of the reaction rates associated with elements of $\mathcal{R}$. 

\begin{figure}[p]
\centering
\includegraphics[width=\textwidth]{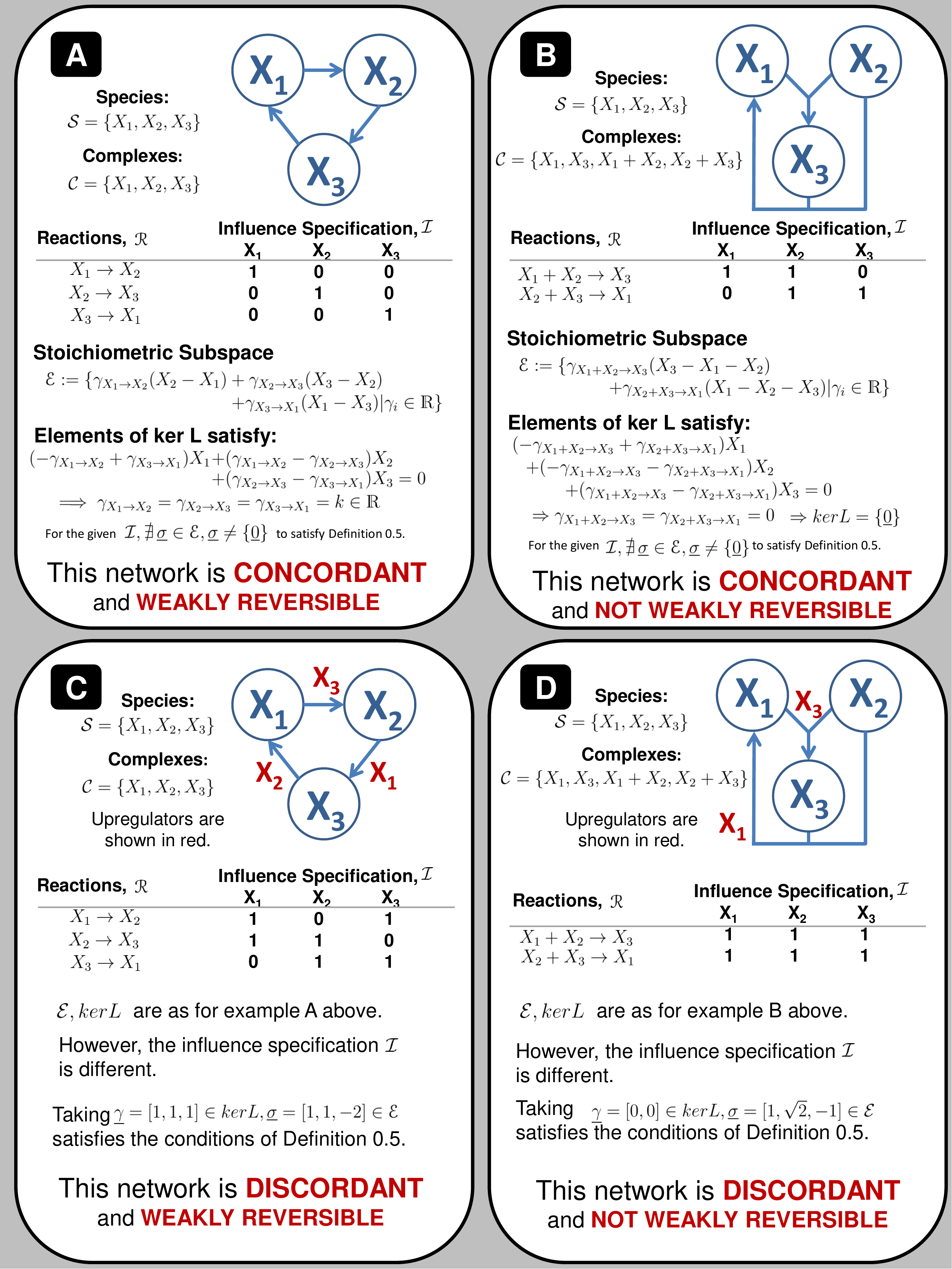}
\caption[]{Simple examples of CRNT network analysis, indicating the sets $\mathcal{S, C, R, I}$ and $\mathcal{E}$. In pairwise cases (A,C) and (B,D), it is the change of influence specification, where new upregulators are introduced, which changes the concordant networks to discordant ones.}
\label{fig: CRNT-example}
\end{figure}

Fig.~\ref{fig: CRNT-example} demonstrates four example networks: (A) concordant and weakly reversible, (B) concordant and not weakly reversible, (C) discordant and weakly reversible, and (D) discordant and not weakly reversible.

\paragraph*{Conventions and Definitions}
We begin by establishing formal definitions for the \emph{influence specification}, $\mathcal{I}$, and \emph{stoichiometric subspace}, $\mathcal{E}$, of a reaction network, and for the properties of \emph{injectivity} and \emph{concordance}. All definitions adopt the following notation conventions:
\begin{itemize}
 \item $\mathbb{R}^I$, where $I$ is a set, denotes the vector space of real-valued functions with domain $I$. This removes the restriction of enumerating vector entries as in $\mathbb{R}^N$ (for $N \in \mathbb{N}$); instead vector entries are indexed over the elements of $I$;
 \item Those vector functions which take only positive values form the subset \mbox{$\mathbb{R}^I_+ \subset \mathbb{R}^I$}; those which take non-negative values are indicated by the set $\overline{\mathbb{R}}^I_+ \subset \mathbb{R}^I$;
 \item The \emph{support} of $\underline{x} \in \mathbb{R}^I$, $supp(\underline{x})$, is the set of indices $i \in I$ for which $x_i \neq 0$;
 \item For $y \in \mathbb{R}$, $sgn(y)$ denotes the sign of $y$; for $y \in \mathbb{R}^I$, $sgn(\underline{y})$ indicates the function such that $(sgn(\underline{y}))_i := sgn(y_i) \,, \forall i \in I$.
\end{itemize}

Having established these notation conventions, we may proceed to some formal definitions. Key definitions are demonstrated for simple networks in Figs. \ref{fig: CRNT-example}.

\begin{framed}
\begin{mydef} \label{def: influence-specification}
An influence specification $\mathcal{I}$ for a reaction network $\{\mathcal{S,C,R}\}$ is an assignment to each reaction $y \rightarrow y'$ of a function \mbox{$\mathcal{I}_{y \rightarrow y'}: \mathcal{S} \rightarrow \{1,0,-1\}$} such that
\begin{itemize}
 \item $\mathcal{I}_{y \rightarrow y'}(s) = 1, \quad \forall s \in supp(y)$,
 \item if $\mathcal{I}_{y \rightarrow y'}(s) = 1$ [resp. $-1$], then species $s$ is an inducer [inhibitor] of reaction $y \rightarrow y'$;
 \item if $\mathcal{I}_{y \rightarrow y'}(s) = 0$, then species $s$ has no influence on the rate of the reaction $y \rightarrow y'$.
\end{itemize}
\end{mydef}
\end{framed}
The \emph{influence specification} therefore assigns each species a $1$, $0$ or $-1$ for every reaction in a network, according to whether the species is an inducer, neutral or an inhibitor of a given reaction. This construct permits the study of networks involving autoregulatory components and allows reactions to be modulated by species other than their products or reactants.

\begin{framed}
\begin{mydef} \label{def: stoi-subspace}
The stoichiometric subspace $\mathcal{E}$ of a reaction network $\{\mathcal{S,C,R}\}$ is the linear subspace of $\mathbb{R}^\mathcal{S}$ defined by
\begin{align*}
 \mathcal{E} := span\{y' - y \in \mathbb{R}^\mathcal{S} |\, y \rightarrow y' \in \mathcal{R}\}.
\end{align*}
Two vectors $\underline{c},\underline{c}^* \in \overline{\mathbb{R}}^\mathcal{S}_+$ are said to be stoichiometrically compatible if $\underline{c}^* - \underline{c} \in \mathcal{E}$.
\end{mydef}
\end{framed}
It is clear from Definition \ref{def: stoi-subspace} that elements of $\mathcal{E}$ are summations of the linear expressions which arise from rearranging reaction expressions to the form $(products - reactants)$. We also define a linear map $L:\mathbb{R}^\mathcal{R} \rightarrow \mathcal{E}$ by
\begin{align}
 L\underline{\gamma} := \sum_{y\rightarrow y' \in \mathcal{R}} \gamma_{y \rightarrow y'} (y' - y)\,, \label{eqn: alpha-transform}
\end{align}
for $\gamma_{y \rightarrow y'} \in \mathbb{R}$ the entries of $\underline{\gamma}$. The map L is used in Definition \ref{def: concordance} to outline concordance. The kernel of $L$ is the set $ker L = \{ \underline{x} \in \mathbb{R}^\mathcal{R} : L(\underline{x}) = \underline{0}\}$.

All elements of $\mathcal{E}$ can be expressed in the form (\ref{eqn: alpha-transform}). However, the standard approach is to write $\underline{\sigma} \in \mathcal{E}$ as a vector in $\mathbb{R}^\mathcal{S}$.

\begin{framed}
\begin{mydef} \label{def: weakly-monotonic-kinetics}
A kinetics $\mathcal{K}$ for a reaction network $\{\mathcal{S,C,R}\}$ is weakly monotonic with respect to influence specification $\mathcal{I}$ if, for every pair of elements $\underline{c}^*, \underline{c}^{**} \in \overline{\mathbb{R}}^\mathcal{S}_+$, the following implications hold for each reaction $y \rightarrow y' \in \mathcal{R}$ such that $supp(y) \subset supp(\underline{c}^{*})$ and \mbox{$supp(y) \subset supp(\underline{c}^{**})$}:
\begin{itemize}
 \item $\mathcal{K}_{y \rightarrow y'}(\underline{c}^{**}) > \mathcal{K}_{y \rightarrow y'}(\underline{c}^{*}) \implies \exists$ species $s$ such that \\ $sgn(c^{**}_s - c^{*}_s) = \mathcal{I}_{y \rightarrow y'}(s) \neq 0$ \,,
 \item $\mathcal{K}_{y \rightarrow y'}(\underline{c}^{**}) = \mathcal{K}_{y \rightarrow y'}(\underline{c}^{*}) \implies$ either: \\
 (a) $c^{**}_s = c^{*}_s \quad \forall s \in supp(y)$, or: \\ 
 (b) $\exists$ species $s, s'$ with $sgn(c^{**}_s - c^{*}_s) = \mathcal{I}_{y \rightarrow y'}(s) \neq 0$ \mbox{and $sgn(c^{**}_{s'} - c^{*}_{s'}) = -\mathcal{I}_{y \rightarrow y'}(s') \neq 0$}.
\end{itemize}
\end{mydef}
\end{framed}
\emph{Weakly monotonic kinetics} therefore admit - amongst others - Hill kinetics, mass-action kinetics and hyperbola functions. Our Notch-Wnt ODE model satisfies the conditions for weakly monotonic kinetics subject to its influence specification $\mathcal{I}$.

\begin{framed}
\begin{mydef} \label{def: injectivity}
A kinetic system $\mathcal{K}$ is injective if, for each pair of distinct, stoichiometrically compatible elements $\underline{c}^{*}, \underline{c}^{**} \in \overline{\mathbb{R}}^\mathcal{S}_+$, at least one of which is positive,
\begin{align*}
 \sum_{y \rightarrow y' \in \mathcal{R}} \mathcal{K}_{y \rightarrow y'}(\underline{c}^{**})(y' - y) \neq \sum_{y \rightarrow y' \in \mathcal{R}} \mathcal{K}_{y \rightarrow y'}(\underline{c}^{*})(y' - y).
\end{align*}
\end{mydef}
\end{framed}
\begin{myremark} \label{rem: monostability}
An injective kinetic system cannot admit two distinct, stoichiometrically compatible equilibria, at least one of which is positive. That is, injectivity may be equated with at most one positive equilibrium.
\end{myremark}

The following definition of concordance relies upon the linear mapping $L$, detailed in Eqn. (\ref{eqn: alpha-transform}).
\begin{framed}
\begin{mydef} \label{def: concordance}
A reaction network $\{\mathcal{S,C,R}\}$ with stoichiometric subspace $\mathcal{E}$ is concordant with respect to influence specification $\mathcal{I}$ if there do not exist $\underline{\gamma} \in ker L$ and a non-zero $\underline{\sigma} \in \mathcal{E}$ having the following properties:
\begin{itemize}
 \item For each $y \rightarrow y'$ such that $\gamma_{y \rightarrow y'} > 0$, there exists a species $s$ for which $sgn(\sigma_s) = \mathcal{I}_{y \rightarrow y'}(s) \neq 0$;
 \item For each $y \rightarrow y'$ such that $\gamma_{y \rightarrow y'} < 0$, there exists a species $s$ for which $sgn(\sigma_s) = -\mathcal{I}_{y \rightarrow y'}(s) \neq 0$;
 \item For each $y \rightarrow y'$ such that $\gamma_{y \rightarrow y'} = 0$, either:\\ 
 (a) $\sigma_s = 0 \quad \forall s \in supp(y)$, or:\\
 (b) $\exists$ species $s, s'$ for which $sgn(\sigma_s) = \mathcal{I}_{y \rightarrow y'}(s) \neq 0$ and \mbox{$sgn(\sigma_s') = -\mathcal{I}_{y \rightarrow y'}(s') \neq 0$};
\end{itemize}
\end{mydef}
\end{framed}
Fig.~\ref{fig: CRNT-example} depicts four simple networks and expressions for $\mathcal{E}$ and $ker L$. The examples in Figs. \ref{fig: CRNT-example}B and \ref{fig: CRNT-example}D have trivial kernels; differing influence specifications mean that the first is concordant according to Definition \ref{def: concordance}, while the other is discordant. Furthermore, for the network in Fig.~\ref{fig: CRNT-example}C, a non-zero $\underline{\sigma} \in S$ and $\underline{\gamma} \in ker L$ can be found which satisfy the conditions listed in Definition \ref{def: concordance}; this is not the case for Fig.~\ref{fig: CRNT-example}A. The networks shown in Figs. \ref{fig: CRNT-example}C and \ref{fig: CRNT-example}D are therefore discordant. None of the examples need be allied with specific reaction rates during this analysis; concordance is a property of the underlying network $\{\mathcal{S,C,R}\}$ and is independent of the kinetics $\mathcal{K}$.

Our definitions now established, we turn to the main theoretical result of interest.

\paragraph*{Concordant Networks with Weakly Monotonic Kinetics}
The following proposition provides us with a means of determining when our Notch and Wnt networks (or indeed the full coupled system) are monostable.
\begin{framed}
\begin{myprop}
\label{prop: injectivity}
A kinetic system $\{\mathcal{S,C,R,K}\}$ is injective whenever there exists an influence specification $\mathcal{I}$ such that:
\begin{itemize}
 \item The kinetics $\mathcal{K}$ is weakly monotonic with respect to $\mathcal{I}$;
 \item The underlying network $\{\mathcal{S,C,R}\}$ is concordant with respect to $\mathcal{I}$.
\end{itemize}
\end{myprop}
\end{framed}

Our Notch-Wnt system has a valid influence specification according to Definition \ref{def: influence-specification} and satisfies Definition \ref{def: weakly-monotonic-kinetics} of weakly monotonic kinetics. Proposition \ref{prop: injectivity} and Remark \ref{rem: monostability} together imply that any concordant network or sub-network in our model should be injective and hence monostable. Conversely, a discordant network or sub-network will exhibit more than one steady state and may exhibit nontrivial dynamics.

\subsection*{Parametrisation Details}
The following details supplement the description in \emph{Methods and Models - Parametrisation} of the main text.

Parameter estimates resulting from computational fitting are assumed to represent a population average; the non-compartmental nature of our model assumes that reactant species are present at a uniform concentration throughout the cell.

The Notch parameter set was tested in a heterogeneous, two-cell system to determine the range of behaviours of the oscillation period as the initial conditions for either cell are varied independently over the range $[0.0, 1.0]$. Exponents $m_i,n_i$ for the Hill functions and hyperbolas are set at $m_i = 3$, for $i = 1,2,4,6,7$ and $n_i = 3$, for $i = 2,3,5$. This reflects the strength of feedback required to generate oscillations in the absence of a delay-driven formulation; exponents are fixed at these stated values and are not included in the parameter fitting exercise. Parameter fitting was performed using \matlab: each parameter was modified in turn over a $\pm100\%$ tolerance (this aims to keep parameter values close to the desired order of magnitude, for biological plausibility); the ODE model was simulated for $1000$ evenly spaced values within this parameter range and the mean period of the oscillations was measured for $12$ hours after stimulation (the oscillation period was defined as the difference between two successive local maxima on the \hes{} timeseries vector). In each case, the parameter estimate was revised to the value which provided the closest match to the two hour oscillation period observed by Hirata \etal{} \cite{hirata2002oscillatory}. This revised value was accepted into the parameter set and the fitting algorithm then moved to the next parameter in the priority set.

The surface plot in Fig.~\ref{fig: hirata-matching} depicts the variation in the oscillation period of the first cell. Overestimation is to be expected for a non-delay model of this kind; the inclusion of Hes1 mRNA and dimerisation processes has been shown to improve matching to oscillatory experimental data \cite{monk2003oscillatory, momiji2008dissecting}. Such modifications might provide the basis for more extensive investigation in the future but lie outside the scope of the present paper.

We now turn to parametrisation of the Wnt system. The Hern\'andez \etal{} timecourse of Fig.~\ref{fig: hernandez-matching} is characterised by a transient accumulation phase for \bcat{} in the two hours post Wnt stimulation, followed by a plateau phase of approximately four hours. During this time, the concentration of the destruction complex remains approximately constant. 

We generate initial estimates of $\alpha_3$ and $\alpha_4$ as follows. From the data in Fig.~\ref{fig: hernandez-matching} we estimate $B \approx 9$ and $\frac{dB}{dt} \approx 0.5$ at $t \sim 0$, and $B \approx 54$ and $\frac{dB}{dt} \approx 0.0$ at $t \sim 400$. If we substitute these values in Eqn. (\ref{eqn: bcat}), assuming further that $\alpha_1 F + \mu_B \approx 0.007$ and fixing $W = 1.0$, then we obtain simultaneous equations for $\alpha_3, \alpha_4$:
\begin{align}
 \frac{dB}{dt} &= (1 + W)\alpha_4 - B(\alpha_1 \cdot F + \alpha_3 \cdot C + \mu_B)\,, \nonumber \\
 0 &= 2\alpha_4 - 54(40\alpha_3 + 0.007)\,, \label{eqn: hernandez-plateau} \\
 0.5 &= 2\alpha_4 - 9(40\alpha_3 + 0.007)\,. \label{eqn: hernandez-slope}
\end{align}
In deriving estimates for $\alpha_3, \alpha_4$ from the experimental data of Hern\'andez \etal{} \cite{hernandez2012kinetic}, we interpret their Wnt concentration as the reference state and set this to be $W=1$ in our model. The substitution $\alpha_1 F + \mu_B \approx 0.007 min^{-1}$ is based upon typical values from the decoupled Notch system at steady state. The steady state concentration of $C$ is estimated using data from Tan \etal{} \cite{tan2012wnt}. The solution of Eqns. (\ref{eqn: hernandez-plateau}) and (\ref{eqn: hernandez-slope}) yields initial estimates of  \mbox{$\alpha_3 = 1.028 \times 10^{-4}$} and \mbox{$\alpha_4 = 0.3$} for use in parameter fitting. Parameter fitting was performed using \matlab. Each parameter in turn was modified over a $\pm100\%$ tolerance (again, to keep parameter values close to the desired order of magnitude); the ODE model was simulated for $1000$ evenly spaced values within this parameter range, before being corrected to the value which minimised the mean-squared error. The evolution of \bcat{} concentration in our parametrised Wnt system is shown in Fig.~\ref{fig: hernandez-matching}.

\subsection*{Sensitivity Analysis}
Parameter sensitivity analyses were performed to create a priority ordering of the parameters for the fitting procedure, using \matlab's Systems Biology Toolbox. The sensitivity, $S_k$ of $X$, where $X(k)$ is either the \bcat{} steady state $B^*$ or the Hes1 oscillation period $T$, to a given parameter $k$ is defined by the following formula:
\begin{align*}
 S_k = \frac{\left\vert X(k + \delta_k) - X(k)\right\vert}{\delta_k},
\end{align*}
where $\delta_k$ is the incremental change in the parameter $k$, for the given parameter set using parameter $k$. This is converted into the \emph{normalised sensitivity index}, $NS_k$, by
\begin{align*}
 NS_k = \frac{k}{X(k)} \times S_k.
\end{align*}
Normalised sensitivities are used to create a parameter priority ordering, with the most sensitive parameters fitted first.

\begin{figure}[htbp]
\centering
\includegraphics[width=\textwidth]{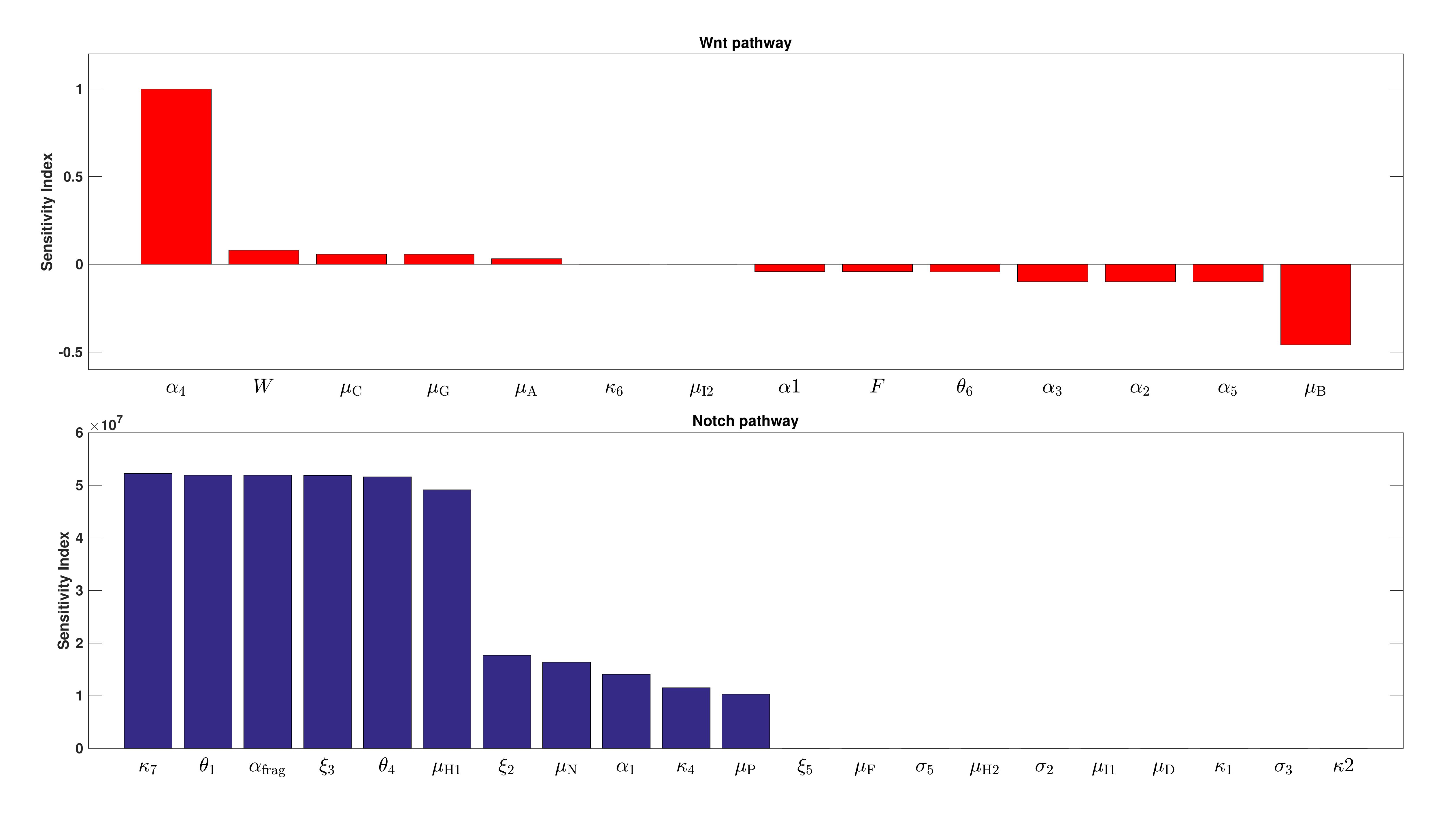}
\caption[]{Results from a preliminary sensitivity analysis for (Top) the steady state of \bcat{} in the Wnt system, (Bottom) the oscillation period of Hes1 in the Notch system, showing normalised sensitivities for a 100\% increase in the value of each parameter in turn. Values for $W$ and $F$ are included here as a comparison, as they are held constant within the decoupled Wnt system; however they are not varied during parameter fitting, as they are variables rather than parameters in the full system. $x1$ is the multiplier for the Wnt response function $\Psi_{W,A}$, described in the Wnt Pathway Submodel of the SI; $\alpha_1$ is fitted using the Notch-only system and is not varied during parameter fitting for the Wnt system.}
\label{fig: Sensitivity}
\end{figure}

Analysis of the decoupled system reveals the most sensitive parameter of the \bcat{} steady state $B^*$ is $\alpha_4$ (a parameter involved in rate of production of \bcat{} subject to Wnt stimulus $W$) whereas the parameter to which period of Hes1 oscillations is most sensitive is $\kappa_7$ (a dissociation rate constant involved in Wnt-mediated transcription of Hes1). A summary of the sensitivity results is presented in Fig.~\ref{fig: Sensitivity}.

%
%
%

%
%

\bibliography{../../References.bib}{}

\end{document}